\documentclass[journal,two column]{IEEEtran}

\usepackage{amsmath,amssymb}
\usepackage{framed}
\usepackage{amsfonts}
\usepackage{graphicx,graphics,color,psfrag}
\usepackage{cite,balance}
\usepackage{caption}
\allowdisplaybreaks
\usepackage{algorithm}
\usepackage{algpseudocode}
\usepackage{accents}
\usepackage{amsthm}
\usepackage{bm}
\usepackage{url}
\usepackage[english]{babel}
\usepackage{multirow}
\usepackage{enumerate}
\usepackage{cases}
\usepackage{stfloats}
\usepackage{dsfont}
\usepackage{color,soul}
\usepackage{amsfonts}
\usepackage{cite,graphicx,amssymb}
\usepackage{fancyhdr}
\usepackage{hhline}
\usepackage{array,color}
\usepackage{mathtools}
\usepackage{subcaption}
\usepackage[most]{tcolorbox}
\newtcolorbox{mybeamdesign}[2][]{
	enhanced,
	colback=white,           % 框内背景颜色：白色
	colframe=black,          % 边框颜色：黑色
	colbacktitle=white,      % 标题栏背景：白色（去掉黑色底色）
	coltitle=red,          % 标题文字颜色：黑色
	center title,
	sharp corners,             % 如果要直角才取消注释，圆角千万别写 =false
	boxrule=1pt,                 % 边框粗细
	left=0pt, right=0pt,       % 内部左右边距
	top=0pt, bottom=0pt,         % 内部上下边距
	title={#2},                  % 标题内容
}

\graphicspath{{./Figs/}}
\newtheorem{definition}{\emph{\underline{Definition}}}

\newtheorem{lemma}{\emph{\underline{Lemma}}}

\newtheorem{example}{\bf \emph{\underline{Example}}}
\newtheorem{remark}{\bf \emph{\underline{Remark}}}

\def\({\left(}
\def\){\right)}

\def\b0{{\mathbf{0}}}

\newcommand{\diag}{\mathrm{diag}}

\makeatletter
\g@addto@macro\normalsize{%
	\setlength\abovedisplayskip{2.7pt}  % 正文到公式上方的距离
	\setlength\belowdisplayskip{2.7pt}  % 正文到公式下方的距离
}
\makeatother

\title{Mitigating Mixed-field Interference in Near-field and Far-field Communications: An Antenna Selection Approach}

\author{Tianyu Liu, Changsheng You,~\IEEEmembership{Member,~IEEE}, Chao Zhou, Mingjiang Wu, \\ Ming-Min Zhao,~\IEEEmembership{Senior Member, IEEE}, and Zhaocheng Wang,~\IEEEmembership{Fellow,~IEEE}
	\vspace{-20pt}
	\thanks{Tianyu Liu, Changsheng You, Chao Zhou and Mingjing Wu are with the Department of Electronic and Electrical Engineering, Southern University of Science and Technology, Shenzhen 518055, China (e-mail: $\!$liuty2022@mail.sustech.edu.cn; youcs@sustech.edu.cn; zhouchao2024@mail.sustech.edu.cn; wumj@sustech.edu.cn).}
	\thanks{Ming-Min Zhao is with the College of Information Science and Electronic Engineering, Zhejiang University, Hangzhou 310027, China, and also with the Zhejiang Provincial Key Laboratory of Multi-Modal Communication Networks and Intelligent Information Processing, Hangzhou 310027, China (e-mail: zmmblack@zju.edu.cn).}
	\thanks{Zhaocheng Wang is with the Department of Electronic Engineering, Tsinghua University, Beijing 100084, China (e-mail: zcwang@tsinghua.edu.cn).}
	\thanks{\emph{Corresponding author: Changsheng You.}}
}

\begin{document}

\captionsetup[figure]{name={Fig.},labelsep=period,singlelinecheck=off} 

\maketitle
\begin{abstract}
	In mixed near-field and far-field systems, the \emph{non-orthogonality} between near-field and far-field channels may cause severe inter-user interference and hence degrade rate performance, when the analog beamforming is designed based on the low-complexity \emph{full-array} maximum ratio transmission (MRT). To tackle this issue, we propose in this paper an \emph{antenna-selection}-based transmission framework to effectively suppress mixed-field interference without mechanically altering antenna structures. To this end, an optimization problem is formulated to maximize the sum-rate of mixed-field systems, by jointly designing antenna selection and power allocation under the MRT-based analog beamforming. As the problem is non-convex and generally difficult to solve optimally, we first consider a typical two-user scenario to obtain useful insights. Interestingly, we analytically show that the strong mixed-field interference can be \emph{substantially} mitigated by deactivating only a \emph{small} portion of antennas, yet without compromising array gains too much. Moreover, an inherent tradeoff is revealed in antenna selection between interference suppression and array-gain enhancement, based on which a suboptimal number of deactivated antennas for achieving the maximum sum-rate is obtained.  Next, for the general multi-user case, we develop an efficient penalty dual decomposition (PDD)-based two-layer framework to obtain its high-quality solution by using the block coordinate descent (BCD) and successive convex approximation (SCA) techniques. To further reduce the computational complexity, a low-complexity antenna-deactivation strategy is proposed capitalizing on an interference-suppression criterion. Last, numerical results demonstrate that the proposed scheme achieves a favorable trade-off between interference suppression and array gain loss, hence achieving significant performance gains over various baseline schemes.
\end{abstract}

\vspace{-20pt}
\section{Introduction}
Extremely large-scale arrays (XL-arrays) have been widely recognized as a key enabling technology to enhance the spectral efficiency and spatial resolution of future wireless communication systems~\cite{You_Changsheng2025tutor}. 
In particular, owing to the continuous expansion of array apertures and the migration of carrier frequencies toward millimeter-wave and even terahertz bands, the Rayleigh distance increases dramatically. As such, a considerable portion of users may be located in the near-field region of the XL-array~\cite{Liu_Yuanwei2023,Cui_Mingyao2023v10,Changsheng_You2023beam,Cong_Jiayi_2024Dec}, necessitating a fundamental paradigm shift in electromagnetic channel modeling, from conventional far-field planar wavefronts (PWs) to the more accurate near-field spherical wavefronts (SWs)~\cite{Cui_Mingyao2023v10}.

This unique channel characteristic enables a new function of \emph{beam-focusing}, which allows to concentrate the beam energy at specific spatial locations/regions~\cite{Zhang_Haodong2026}. This introduces new design opportunities for multi-user transmissions and interference management in wireless systems. Specifically, the authors in \cite{Zhang_Haiyang2022} demonstrated that users sharing the same angular direction can still be simultaneously served if they are located at different ranges by exploiting the beam-focusing gain in the range domain. Moreover, it was also shown that such beam-focusing gains can be preserved under hybrid beamforming architectures. Motivated by these advantages, the design of efficient hybrid beamforming strategies for near-field systems has attracted significant attention. Among others, one commonly adopted approach is by using alternating optimization technique, where digital and analog precoders are iteratively optimized \cite{Liu_Xinhao2025,Zhou_Cong2025bemforming}. However, handling the constant-modulus constraint in the analog domain typically incurs prohibitively high computational complexity as the array size grows.
To improve robustness and tractability, a fully-digital approximation method was proposed in \cite{Wang_Zhaolin2025}, where the analog and digital precoders are designed to approximate the ideal unconstrained fully-digital precoder by minimizing the Frobenius-norm least-squares distance between them.
However, the high-dimensional matrix inversions and projections involved still incur substantial computational complexity for XL-array systems. To further alleviate this complexity bottleneck, heuristic two-stage beamforming designs were studied in \cite{Wang_Zhaolin2023Conf,Liujia_Yao2026}, where the analog beams are first steered toward user locations to achieve the beamforming gain, followed by a low-dimensional digital precoding to suppress multi-user interference. As a result, two-stage schemes achieve favorable spectral efficiency with significantly reduced computational complexity, offering a practical and scalable solution for near-field communications~\cite{Liu_Yuanwei2023}.

However, in practice, near-field and far-field users may coexist in the same system, leading to complex \emph{mixed-field} channel correlations due to mismatched channel characteristics, which raises new research issues in communication system designs~\cite{Wang_Zhaocheng2025,Chen_Yuanbin2024,Wang_Kuiyu2024}. Specifically, it was revealed in~\cite{Cui_Mingyao2022} that, due to the inherent \emph{non-orthogonality} between near-field and far-field channel steering vectors, a beam steered toward a far-field user can spread its energy over a wide angular regime in the near-field region, a phenomenon commonly referred to as the \emph{energy-spread effect}. On one hand, this effect can be exploited for performance enhancement through appropriate beamforming designs. For example, the authors in~\cite{Zhang_Yunpu2024} leveraged the energy-spread characteristic of mixed-field propagation to enable efficient wireless energy transfer in simultaneous wireless information and power transfer (SWIPT) systems, while the authors in~\cite{Liu_Tianyu2025} demonstrated that mixed-field interference can be harnessed as beneficial interference to enhance physical-layer security. On the other hand, the energy-spread effect may cause severe inter-user interference, where near-field (or far-field) users experience strong power leakage from far-field (or near-field) beams over a broad angular regime, as revealed in~\cite{YZhang2023}. 
To mitigate such mixed-field interference, several approaches have been recently proposed by exploiting reconfigurable antennas. For example, rotatable-antenna technique was studied in~\cite{Yunpu_Zhang2025} to reduce near-far channel correlation by dynamically adjusting antenna orientations, while movable-antenna framework was investigated in~\cite{Chao_Zhou2025} to suppress near-far signal leakage and enhance transmission security through antenna position reconfiguration. Nevertheless, these approaches generally rely on mechanically reconfigurable antenna structures, which inevitably increase hardware complexity and implementation cost.  Therefore, it is of great importance to investigate how to deal with near-far channel correlation for improving mixed-field communication performance without mechanically altering antenna structures (e.g., no mechanical antenna movement/rotation). 
\begin{figure}[t]
	\centering
	\includegraphics[width=3.5in]{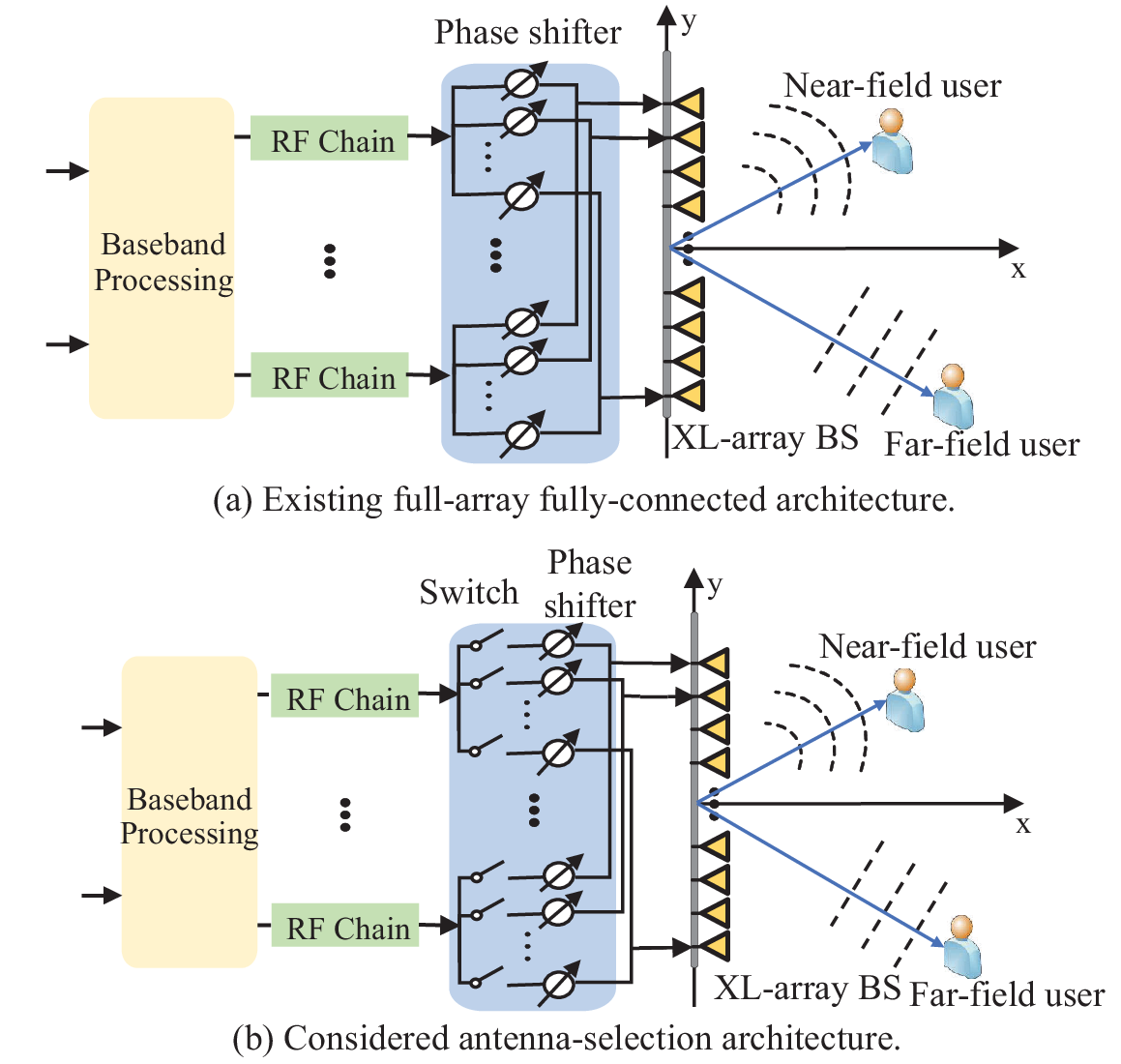}%{system_model_mixed_field.eps}
	\caption{Different beamforming architectures for the considered mixed-field communication system.}
	\label{system_setting}
\end{figure}

\vspace{-10pt}
\subsection{Motivations and Contributions}
Motivated by the above, we consider an \emph{antenna-selection}-based beamforming framework in this paper for mixed-field XL-array systems, as shown in Fig.~\ref{system_setting}(b). To reduce implementation complexity, a two-stage hybrid beamforming architecture is considered where the analog beamforming is designed based on MRT. Note that when the far-field user is located within a certain angular regime close to the near-field user, there may exist strong interference between them. To tackle this issue, by integrating a switch network into the conventional hybrid beamforming architecture and judiciously activating a subset of antennas, antenna selection can \emph{reshape the effective array structure} and introduce additional spatial degrees-of-freedom~\cite{Nai_SiewEng2010, Wang_Xiangrong2014}. This enables improved control over \emph{mixed-field channel correlations} and provides an effective means to suppress mixed-field inter-user interference. The main contributions of this paper are summarized as follows.
\begin{itemize} 
	\item First, we consider a typical two-user mixed-field scenario to reveal the fundamental role of antenna selection in promoting the near-field and far-field channel \emph{orthogonality}. A low-complexity antenna deactivation algorithm is proposed based on an interference-suppression criterion. In particular, we show that the severe mixed-field interference induced by full-array MRT beams can
	be reduced linearly without incurring a significant array-gain loss, where the interference decreasing rate is approximately $1/\sqrt{N}$ with $N$ denoting the total number of antennas. Furthermore, we characterize the inherent trade-off between interference suppression and array-gain preservation in antenna selection, based on which, a suboptimal solution of the number of deactivated antennas for achieving the maximum sum-rate is obtained.
	\item Second, for the general multi-user case, we formulate a sum-rate maximization problem by jointly designing the antenna selection and power allocation under the MRT-based analog beamforming. The formulated problem is non-convex and generally difficult to optimally solve. To tackle this difficulty, we first reformulate it into an augmented Lagrangian form and then develop an efficient penalty dual decomposition (PDD)-based two-layer optimization framework. Specifically, the inner layer efficiently solves the problem via block coordinate descent (BCD), while the outer layer updates the dual variables and penalty parameter. To further reduce the computational complexity, a low-complexity yet efficient antenna deactivation method is developed by first designing antenna selection based on the multi-user quasi-in-phase antenna deactivation, followed by the second phase of power allocation.
\item Last, numerical results are presented to demonstrate the performance gains of the proposed PDD-based and low-complexity schemes over various baselines. Specifically, the gains over different baselines validate the effectiveness of antenna selection in mitigating mixed-field interference. Furthermore, the superiority over different baseline architectures demonstrates that our proposed flexible design strikes a superior balance between array gain preservation and interference suppression. Moreover, the proposed low-complexity deactivation scheme achieves performance close to that of the PDD-based solution with substantially reduced computational complexity, while maintaining robustness under multi-path propagation and channel estimation errors.
\end{itemize}

\vspace{-10pt}
\subsection{Organization and Notations}
The remainder of this paper is organized as follows. Section II introduces the system model for mixed-field communications. Then we consider an antenna-selection-based architecture in Section III, where a sum-rate maximization problem is formulated for general multi-user scenarios. To shed further insights, Section IV focuses on a typical two-user mixed-field scenario and analyzes the performance gains achieved by antenna selection. In Section V, we develop an efficient algorithm to obtain a high-quality suboptimal solution for the general case and further propose a low-complexity antenna deactivation strategy. Numerical results are presented in Section VI to evaluate the effectiveness of the proposed schemes, followed by conclusions drawn in Section VII.

\emph{Notations:}  Lowercase and uppercase boldface letters are
used to represent vectors and matrices, e.g., $\mathbf{a}$ and $\mathbf{A}$. The notation $[\mathbf{a}]_n$ denotes the $n$-th entry of vector $\mathbf{a}$, and $[\mathbf{A}]_{n,n}$ denotes the $n$-th diagonal element of matrix $\mathbf{A}$. The superscripts $(\cdot)^T$, $(\cdot)^*$, and  $(\cdot)^H$ stand for the transpose, Hermitian and Hermitian transpose, respectively. Additionally, calligraphic letters (e.g., $\mathcal{A}$) denote discrete and finite sets. The notations $|\cdot|$,  $\|\cdot\|$, and $\|\cdot\|_F$ refer to the absolute value of a scalar, the $\ell_2$ norm of  a vector, and the Frobenius norm of a matrix, respectively. The symbol $\mathbf{I}_K$ represents a $K$-dimensional identity matrix. $\mathcal{O}(\cdot)$ denotes the standard big-O notation.

\vspace{-5pt}
\section{System Model}
We consider a downlink mixed-field communication system as shown in Fig.~\ref{system_setting}, where a BS equipped with an $N$-antenna XL-array serves $K$ single-antenna users. The array is placed along the $y$-axis with its center at the origin $(0,0)$, and the $n$-th antenna is located at $(0, \delta_n d)$ in the Cartesian coordinate system, where $\delta_n=(2n-N-1)/2, n \in \mathcal{N}\triangleq\{1,\cdots,N\}$ and $d=\lambda/2$ is the half-wavelength inter-antenna spacing. The boundary between near-field and far-field regions is given by the effective Rayleigh distance $Z(\theta)=2\epsilon D^2(1-\theta^2)/\lambda$, where $D = (N-1)d$ is the array aperture, $\theta \in [-1, 1]$ is the spatial angle, and $\epsilon=0.367$~\cite{Cui_Mingyao2024}. Note that the effective Rayleigh distance is considered in this paper, which is directly related to the array gain and hence the communication performance~\cite{Cui_Mingyao2024}. Among the $K$ users, $K_1$ users are assumed to be located in the near-field region (denoted by the set $\mathcal{K}_1\triangleq\{1,2,\cdots, K_1\}$ without loss of generality) and the remaining $K_2$ users are in the far-field region (denoted by the set $\mathcal{K}_2\triangleq\{K_1+1,\cdots, K\}$).
Furthermore, the channel state information (CSI) for both near-field and far-field users is assumed to be available at the BS, by using existing channel estimation methods (see e.g., \cite{Alkhateeb_Ahmed2014, Cui_Mingyao2021estimation, Wu_Chenyu2024,Zhang_Yunpu2022}).\footnote{Specifically, the CSI of far-field users can be acquired via orthogonal matching pursuit (OMP) or multiple signal classification (MUSIC), while that of near-field users can be obtained through polar-domain OMP or various beam training methods \cite{Alkhateeb_Ahmed2014,Schmidt_R1986,Cui_Mingyao2021estimation,Wu_Chenyu2024}. The impact of imperfect CSI will be evaluated in Section~VI, while the robust antenna selection design to mitigate CSI errors is left for future research~\cite{Chao_Zhou2026}.}

For the near-field user $k \in \mathcal{K}_1$, its channel from the BS can be modeled as 
\begin{align}
	\mathbf{h}^H_k &=\mathbf{h}^H_{{\rm LoS},k}+\sum_{q_k=1}^{Q_k}\mathbf{h}^H_{{\rm NLoS},m_k}, \forall k \in \mathcal{K}_1, \label{channelmodel}
\end{align}
which consists of one line-of-sight (LoS) path and $Q_k$ non-LoS (NLoS) paths.
We consider communication systems operating in high-frequency bands, where the power of NLoS paths is negligible due to the severe path-loss and shadowing. As such, the channel of near-field user $k$ can be approximated by its LoS component as\footnote{It is worth noting that the proposed design can be directly applied to the general multi-path channel model. The corresponding performance evaluation under multi-path channels will be presented in Section VI.} 
\begin{align}
		\mathbf{h}^H_k \approx \mathbf{h}^H_{{\rm LoS},k} =\sqrt{N}h_k\mathbf{b}^H(\theta_k,r_k), \forall k \in \mathcal{K}_1, \label{channelmodelLoS}
\end{align}
where $h_k=\frac{\sqrt{\beta}}{r_k}e^{-\frac{j2\pi r_k}{\lambda}}$ is the complex-valued channel gain, with $\beta$ and $r_k$ representing the reference channel gain at a distance of 1 meter (m) and the distance between the BS array center and user $k$, respectively; and  $\theta_k \triangleq \sin{\phi_k}$ denotes the spatial angle at the BS with $\phi_k \in (-\frac{\pi}{2},\frac{\pi}{2})$ representing the physical angle-of-departure (AoD) from the BS array center to user~$k$. Additionally, $\mathbf{b}^H(\theta_k,r_k)$ denotes the near-field channel steering vector, which is modeled as follows based on uniform SWs (USWs)~\cite{Wu_Xun2024}
\begin{align}
	&\mathbf{b}^H\!\!\left(\theta_k,\!r_k\right)\!=\!\frac{1}{\sqrt{N}}\!\left[e^{-j\frac{2\pi}{\lambda}(r^{(0)}_k\!-\!r_k)},\dots,e^{-j\frac{2\pi}{\lambda}(r^{(\!N-1\!)}_k\!-\!r_k\!)}\right]\!,
\end{align}
with $r^{(n)}_k=\sqrt{r^2_k+\delta^2_nd^2-2r_k\theta_k\delta_n d}$ representing the distance between the $n$-th BS antenna and user $k$.

Similarly, for each far-field user $k \in \mathcal{K}_2$, its channel from the BS can also be approximated by its LoS component as
\begin{align}
	\mathbf{h}^H_k \approx \mathbf{h}^H_{{\rm LoS},k} = \sqrt{N}h_k\mathbf{a}^H(\theta_k),  k \in \mathcal{K}_2.
\end{align}
Herein, $\mathbf{a}^H(\theta_k)$ denotes the far-field channel steering vector, which is given by
\begin{align}
	\mathbf{a}^H(\theta_k)=\frac{1}{\sqrt{N}}\left[1, e^{j \pi \theta_k}, \dots, e^{j \pi (N-1) \theta_k} \right].
\end{align}

%\vspace{-5pt}
\section{Considered Hardware Architecture and Problem Formulation}
In this section, we first present the conventional full-array fully-connected hybrid beamforming architecture and discuss its limitations in mixed-field communication systems. To address this issue, we consider the antenna-selection-based beamforming architecture and formulate a sum-rate maximization problem.

\vspace{-5pt}
\subsection{Existing Hybrid Beamforming Architecture}
For the conventional full-array fully-connected hybrid beamforming architecture shown in Fig. 1(a), each of the $K$ RF chains is connected to all antennas through a network of phase shifters (PSs). The transmitted signal is precoded by a digital precoder $\mathbf{F}_{\rm D} = [\mathbf{f}_{{\rm D},1}, \dots, \mathbf{f}_{{\rm D},K}] \in \mathbb{C}^{K \times K}$ followed by an analog precoder $\mathbf{F}_{\rm A} = [\mathbf{w}_1, \dots, \mathbf{w}_K] \in \mathbb{C}^{N \times K}$. The received signal at user $k$ is expressed as $y_k = \mathbf{h}^H_k \mathbf{F}_{\rm A} \mathbf{F}_{\rm D} \mathbf{x} + z_k$, where $\mathbf{x}=[x_1, x_2, \ldots, x_K]^T$ is the signal vector and $z_k \sim \mathcal{CN}(0, \sigma^2)$ is the received noise at user $k$. In conventional hybrid beamforming, the joint design of $\mathbf{F}_{\rm A}$ and $\mathbf{F}_{\rm D}$ generally requires high-dimensional matrix inversions for digital precoding, and sophisticated processing to address the constant-modulus constraints of analog phase shifters, resulting in prohibitively high computational complexity~\cite{ZhiQuan_Luo2006,He_Xin2022}.

To facilitate practical implementation of XL-array, a two-stage hybrid beamforming scheme is considered~\cite{Zhang_Yunpu2024, Liu_Tianyu2025}. Specifically, the analog beamformer is designed as $\mathbf{w}_k=\sqrt{N}\mathbf{b}(\theta_k,r_k), k \in \mathcal{K}_1$ for near-field users and $\mathbf{w}_k=\sqrt{N}\mathbf{a}(\theta_k), k \in \mathcal{K}_2$ for far-field users, aiming to maximize the received signal powers at individual users. Then, to reduce the hardware cost and signal processing complexity of
the XL-array, the digital beamforming $\mathbf{F}_{\rm D}$ is simplified as a diagonal power allocation matrix, i.e., $\mathbf{F}_{\rm D} = \diag\{\sqrt{P_1/N},\sqrt{P_2/N},\cdots,\sqrt{P_K/N}\}$, where $P_k$ is the transmit power allocated to user $k$~\cite{Zhang_Yunpu2024}.\footnote{For XL-array systems, general digital precoding incurs very high complexity due to high-dimensional matrix operations. More importantly, to isolate and highlight the impact of antenna selection on the array geometry and the resulting mixed-field channel correlation, an identity digital precoder is adopted.} Based on the above, the received signal at user $k$ can be rewritten as
\begin{align}
	y_k\!=\!\sqrt{\frac{P_k}{N}}\mathbf{h}^H_k\mathbf{w}_k x_k\!+\!\!\!\sum^K_{i=1,i\neq k}\!\!\sqrt{\frac{P_i}{N}}\mathbf{h}^H_k\mathbf{w}_i x_i\!+\!z_k, \forall k\in\! \mathcal{K},
\end{align} 
where $\mathcal{K}=\mathcal{K}_1 \cup \mathcal{K}_2$ is the set of all users, and $z_k \sim \mathcal{CN}(0, \sigma^2)$ represents the received noise at user $k$.
Accordingly, the achievable rate of user $k$, in bits per second per Hertz (bps/Hz), is given by
\begin{align}
	&R_k=\log _2\!\!\left(1+\frac{\frac{P_k}{N}\left|\mathbf{h}^H_k\mathbf{w}_k\right|^2}{\sum^K_{i=1,i \neq k}\frac{P_i}{N}\left|\mathbf{h}^H_k\mathbf{w}_i\right|^2+\sigma^2}\right)\!,\forall k\in \mathcal{K}. \label{AchinforRate_noselec}
\end{align} 
However, under the full-array fully-connected hardware architecture, inter-user interference can become very severe when the near-field and far-field channels exhibit \emph{high correlation} over the full array, which results in significant rate performance degradation, as illustrated in the following example. 

\begin{figure}[t]
	\centering
	\begin{subfigure}[b]{0.49\linewidth}
		\includegraphics[width=1.1\linewidth]{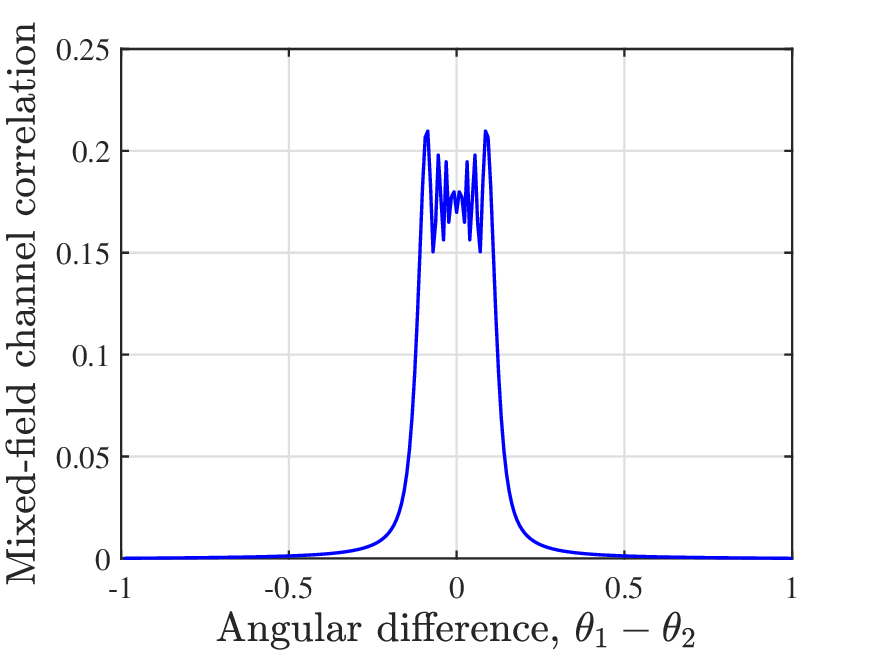}
		\caption{Channel correlation.}
		\label{3D_Illustration}
	\end{subfigure}
	\begin{subfigure}[b]{0.49\linewidth}
		\includegraphics[width=1.1\linewidth]{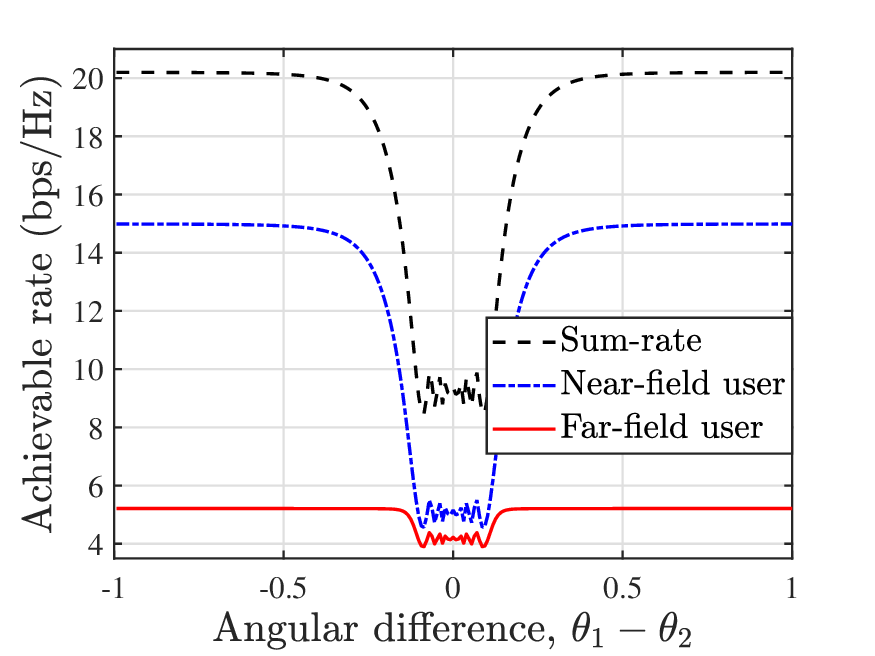}
		\caption{Achievable sum-rate.}
		\label{1D_Illustration}
	\end{subfigure}
	\caption{Mixed-field channel correlation and achievable sum-rate vs. angular difference in the two-user case.}
	\label{G_vs_beta1}
\end{figure}

\begin{example} \rm We consider a mixed-field scenario with one near-field user (i.e., $\mathcal{K}_1=\{1\}$) and one far-field user (i.e., $\mathcal{K}_2=\{2\}$), located at distances of 5~m and 150~m, respectively. We show in Fig.~\ref{G_vs_beta1} the mixed-field channel correlation, i.e., $|\mathbf{b}^H(\theta_1,r_1)\mathbf{a}(\theta_2)|$, as well as the achievable sum-rate versus (vs.) the angular difference between the two users under equal power allocation. One can observe from Fig.~\ref{G_vs_beta1}(a) that the mixed-field channel correlation remains relatively high over a wide angular region due to the so-called \emph{energy-spread} effect~\cite{YZhang2023}, resulting in severe interference and hence degraded rate performance (see Fig.~\ref{G_vs_beta1}(b)).
\end{example}

\vspace{-15pt}
\subsection{Considered Hardware Architecture}  % for flexible beamforming design
%还要写天线如何删减的

To tackle the above issue, we consider an antenna-selection-based hybrid architecture for mixed-field communication systems, as shown in Fig.~\ref{system_setting}(b). Specifically, a switch network is employed to dynamically connect each of the $K$ RF chains to a subset of PSs, which enables adaptive antenna selection and flexible beamforming. Let $\mathbf{V}_k$ denote the diagonal antenna-selection matrix for user $k$, where $[\mathbf{V}_k]_{n,n}  = 1$ if the $n$-th antenna is activated, and $[\mathbf{V}_k]_{n,n}  = 0$ otherwise. The number of activated antennas for user $k$ is $M_k  =\sum_{n=1}^N [\mathbf{V}_k]_{n,n}, \forall k\in \mathcal{K}$. Accordingly, the analog beamforming vector under this architecture can be expressed as $\mathbf{w}^{\rm AS}_k= \mathbf{V}_k \mathbf{w}_k, \forall k\in \mathcal{K}$. 
As such, the received signal at user $k$ can be rewritten as
\begin{align}
	&y_k \!=\! \sqrt{\frac{P_k}{M_k}}\mathbf{h}^H_k\mathbf{w}^{\rm AS}_kx_k\!+\!\!\!\!\sum^K_{i=1,i\neq k}\sqrt{\frac{P_i}{M_i}}\mathbf{h}^H_k\mathbf{w}^{\rm AS}_ix_i\!+\!z_k,
\end{align} 
with its corresponding achievable rate given by
\begin{align}
	R^{{\rm AS}}_{k}\! =\!  \log_2\! \! \Bigg(\! 1\!  +\!  \frac{\frac{P_k}{M_k} |\mathbf{h}_k^H \mathbf{V}_k \mathbf{w}_k|^2}{\sum_{i \neq k} \frac{P_i}{M_i} |\mathbf{h}_k^H \mathbf{V}_i \mathbf{w}_i|^2 + \sigma^2}\!  \Bigg), \! \forall k\in \mathcal{K}. \label{RateUnify}
\end{align}
With MRT-based analog beamforming, the achievable rates in \eqref{RateUnify} for the near-field and far-field users can be respectively rewritten as \eqref{Rate_MRT_NF} and \eqref{Rate_MRT_FF}, shown at the top of the next page.

\begin{figure*}[t] % 强制浮动到页面底部
	\vspace{-2.5em}
	\begin{align}
		\resizebox{0.9\textwidth}{!}{$
			\text{	$R^{{\rm AS}}_{k}\!=\!\log _2\left(1+\frac{ P_k|h_k|^2M_k}{\sum_{i\in \mathcal{K}_2}P_i|h_k|^2\frac{N^2}{M_i}\left|\mathbf{b}^H(\theta_k,r_k)\mathbf{V}_i\mathbf{a}(\theta_i)\right|^2+\sum_{j\in \mathcal{K}_1, j \neq k}P_j|h_k|^2\frac{N^2}{M_j}\left|\mathbf{b}^H(\theta_k,r_k)\mathbf{V}_j\mathbf{b}(\theta_j,r_j)\right|^2+\sigma^2}\right), \forall k \in \mathcal{K}_1,$}
			$}
		\label{Rate_MRT_NF}
		\\
		\resizebox{0.9\textwidth}{!}{$
			\text{$R^{{\rm AS}}_{k}\!=\!\log _2\left(1+\frac{ P_k|h_k|^2M_k}{\sum_{i\in \mathcal{K}_2, i \neq k}P_i|h_k|^2\frac{N^2}{M_i}\left|\mathbf{a}^H(\theta_k)\mathbf{V}_i\mathbf{a}(\theta_i)\right|^2+\sum_{j\in \mathcal{K}_1 }P_j|h_k|^2\frac{N^2}{M_j}\left|\mathbf{a}^H(\theta_k)\mathbf{V}_j\mathbf{b}(\theta_j,r_j)\right|^2+\sigma^2}\right), \forall  k \in \mathcal{K}_2$.}
			$}
		\label{Rate_MRT_FF}
	\end{align}
	\vspace{-1.5em}
			\hrulefill
\end{figure*}

As observed in \eqref{Rate_MRT_NF} and \eqref{Rate_MRT_FF}, antenna selection essentially reshapes the \emph{array configuration}, which introduces additional beamforming design flexibility by judiciously activating a subset of antennas for each user, thereby providing a viable approach to reduce the interferences on the other users. 
\begin{definition}
	\label{def:interference_coupling} \rm
	To quantify the mixed-field interference effect, we define an \emph{interference coupling factor} for each user $k$, denoted by $I_k$, which characterizes the \emph{total interference} imposed by user $k$ on the remaining $(K-1)$ users. Mathematically, $I_k$ is defined as
	\begin{align}
		I_k \triangleq \sum_{\substack{i=1,\ i \neq k}}^{K}
		\frac{1}{\sqrt{M_k}}
		\left|\mathbf{h}_i^H \mathbf{V}_k \mathbf{w}_k\right|,\forall k\in \mathcal{K}.\label{TotInterf}
	\end{align}
\end{definition}

\vspace{-13pt}
\subsection{Problem Formulation} 
Under the antenna-selection-based hardware architecture, we formulate an optimization problem to maximize the achievable sum-rate of all users by jointly optimizing the antenna selection matrices $\{\mathbf{V}_k\}^K_{k=1}$ and the transmit power allocation $\{P_k\}^K_{k=1}$, which is given by\footnote{The proposed framework can be readily extended to the weighted sum-rate problem by incorporating user-specific weighting factors, enabling a flexible fairness-efficiency trade-off as evaluated in Section~VI.}
\begin{align}
	\textbf{(P1)}\;\max_{\{P_k,\mathbf{V}_k\}} \; &\sum^K_{k=1} R^{{\rm AS}}_{k} \nonumber\\
	\text{s.t.}\;\;\;\;\;\;\;\;\;
	&[\mathbf{V}_k]_{n,n}  \! \in \! \{0,1\}, \forall  n \!  \in\!\mathcal{N}\! , \;\forall k \! \in\! \mathcal{K}, \label{P2AntennaSelection}\\
	&\sum^K_{k=1}P_k \leq P_{\rm tot},\label{TotPower}
\end{align}
where $P_{\rm tot}$ denotes the total transmit power. Problem \textbf{(P1)} is a non-convex optimization problem, which is generally difficult to solve optimally, since 1) the optimization variables are tightly coupled in the non-concave objective function; and 2) the binary constraints in \eqref{P2AntennaSelection} render the problem NP-hard. To tackle these difficulties and obtain useful insights, we first focus on a two-user mixed-field scenario, where a low-complexity yet efficient antenna deactivation method is developed to characterize the performance gains achieved by antenna selection. Subsequently, we develop an efficient algorithm to solve Problem \textbf{(P1)}  for the general multi-user scenario.

\vspace{-5pt}
\section{Two-user Case: Interference Analysis and Antenna-selection Design}
To shed important insights into the mixed-field interference suppression capability of antenna selection and validate its performance gains, we focus on a two-user mixed-field scenario in this section, consisting of one near-field user (i.e., $\mathcal{K}_1=\{1\}$) and one far-field user (i.e., $\mathcal{K}_2=\{2\}$). As such, their achievable rates respectively reduce to 
\begin{align}
	&R^{{\rm AS}}_1\!=\!\log _2\!\!\left(\!\!1\!\!+\!\!\frac{P_1|h_1|^2M_1}{P_2|h_1|^2\!\frac{N^2}{M_2}\!\left|\mathbf{b}^H(\theta_1,r_1)\mathbf{V}_2\mathbf{a}(\theta_2)\right|^2\!\!+\!\!\sigma^2}\!\!\right), \label{AchinforRate_selec_uN} \\
	&R^{{\rm AS}}_2\!=\!\log _2\!\!\left(\!\!1\!\!+\!\!\frac{P_2|h_2|^2M_2}{P_1|h_2|^2\!\frac{N^2}{M_1}\!\left|\mathbf{a}^H(\theta_2)\mathbf{V}_1\mathbf{b}(\theta_1,r_1)\right|^2\!\!+\!\!\sigma^2}\!\!\right).   
	\label{AchinforRate_selec_uF}
\end{align}
As such, Problem \textbf{(P1)} reduces to the following optimization problem corresponding to antenna selection
\begin{align}
	\textbf{(P2)}\;\max_{\{P_k,\mathbf{V}_k\}}
	\;R^{{\rm AS}}_1+R^{{\rm AS}}_2
	\qquad
	\text{s.t.} \;\eqref{P2AntennaSelection},\eqref{TotPower}. \nonumber
\end{align}
Problem \textbf{(P2)} can be efficiently solved by applying the BCD method to decouple antenna selection and power allocation. Specifically, for fixed antenna-selection matrices $\{\mathbf{V}_k\}$, the optimal power allocation $\{P_k\}$ can be efficiently obtained via a one-dimensional search. In contrast, the optimal antenna-selection matrices $\{\mathbf{V}_k\}$ can be obtained via exhaustive search or the branch-and-bound method \cite{Shrestha_Sagar2023}, both of which, however, incur exponential computational complexity. To draw useful insight and develop a computationally efficient antenna selection strategy, we first analyze the impact of antenna-selection on the objective function of Problem \textbf{(P2)}, given the power allocation. The key observations are summarized below.
\begin{remark} \label{EffofAnttonRate}\rm The effects of the antenna selection on the sum-rate are summarized as follows.
\begin{itemize}
	\item Activating more antennas for each user can improve its array gain, which is characterized by the number of activated antennas, i.e., $M_k=\sum^N_{n=1} [\mathbf{V}_k]_{n,n}$.
    \item By properly deactivating a portion of antennas, antenna selection can suppress mixed-field interference by reshaping channel correlation, which is characterized by the following interference coupling factors $I_1$ and $I_2$:
\begin{align}
	&I_1=N/\sqrt{M_1}\left|\mathbf{a}^H(\theta_2)\mathbf{V}_1\mathbf{b}(\theta_1,r_1)\right|,\\
	&I_2=N/\sqrt{M_2}\left|\mathbf{b}^H(\theta_1,r_1)\mathbf{V}_2\mathbf{a}(\theta_2)\right|.
\end{align}
\end{itemize}
\end{remark} 
According to \eqref{AchinforRate_selec_uN} and \eqref{AchinforRate_selec_uF} and \textbf{Remark \ref{EffofAnttonRate}}, it can be shown that, given the number of activated antennas, i.e., $M_1$ and $M_2$, maximizing the sum-rate is equivalent to minimizing the inter-user interference, i.e., $I_1$ and $I_2$. As such, given the power allocation, Problem \textbf{(P2)} can be reformulated as a two-layer optimization problem for antenna selection design: the inner problem optimizes the antenna-selection matrix $\mathbf{V}_k$ with $\sum_{n=1}^N[\mathbf{V}_k]_{n,n}=M_k$, while the outer problem optimizes the number of activated antennas $\{M_1,M_2\}$.

%\medskip
\underline{\textbf{Inner problem:}}
Given the number of activated antennas $\{M_1,M_2\}$, the inner problem for antenna-selection matrix optimization is formulated as
\begin{align}
	\textbf{(P2.1)}\quad
	\max_{\{\mathbf{V}_k\}}
	\quad
	&R^{\rm AS}_1+R^{\rm AS}_2 \nonumber\\
	\text{s.t.}\quad
	&\sum_{n=1}^N[\mathbf{V}_k]_{n,n}=M_k, \forall k\in \{1,2\}, \label{TotNumAnn}\\
	&\eqref{P2AntennaSelection}. \nonumber
\end{align}
We denote $	R_{\rm sum}(M_1,M_2) \triangleq R^{\rm AS}_1(M_1,\bar{\mathbf{V}}_2(M_2)) + R^{\rm AS}_2(M_2,\bar{\mathbf{V}}_1(M_1))$, which represents the optimized objective value of Problem \textbf{(P2.1)}, with $\bar{\mathbf{V}}_k(M_k)$ being the obtained antenna-selection matrix.

%\medskip
%\noindent
\underline{\textbf{Outer problem:}}
Given the solution to the inner problem, the outer problem for optimizing the number of activated antennas is given by
\begin{align}
	\textbf{(P2.2)}\quad\max_{M_1,M_2}
	\quad
	&R_{\rm sum}(M_1,M_2)\nonumber\\
	\text{s.t.}
	\quad\;\;
	&M_k \in \{1,\dots,N\},\;  \forall k\in\{1,2\}.
\end{align}

\vspace{-5pt}
\subsection{Inner Problem: Antenna-selection Design for Interference Suppression}

In this subsection, we solve the inner problem \textbf{(P2.1)}
for a given number of activated antennas $\{M_1,M_2\}$.
According to \eqref{AchinforRate_selec_uN} and \eqref{AchinforRate_selec_uF}, when $M_k$ is fixed, maximizing the sum-rate in \textbf{(P2.1)} is equivalent to minimizing the interference coupling factors for both the near-field and far-field users. Thus, Problem \textbf{(P2.1)} can be decoupled into the following two interference-minimization subproblems:
\begin{align}
\text{Near-field user:}\;&\min_{\mathbf{V}_1} \;I_1\;\;\; {\rm s.t.}\; \eqref{P2AntennaSelection},\eqref{TotNumAnn}. \nonumber\\ 
\text{Far-field user:}\;&\min_{\mathbf{V}_2} \;I_2\;\;\;{\rm s.t.}\; \eqref{P2AntennaSelection},\eqref{TotNumAnn}. \nonumber
\end{align}

Without loss of optimality, we focus on the antenna-selection design for the near-field user to minimize its interference to the far-field user (i.e., $I_1$), while the same procedures apply to the far-field user. To this end, we first propose a greedy yet efficient antenna deactivation scheme based on the principle of \emph{quasi-in-phase} antenna deactivation, which is structured into the following three steps:

\textbf{Step 1:} Let $c_n = [\mathbf{a}(\theta_2)]_n^* [\mathbf{b}(\theta_1, r_1)]_n$ denote the contribution of the $n$-th antenna to the spatial correlation. For an initial set of activated antennas $\mathcal{S}^{(0)}=\{1,2,\ldots,N\}$, the interference coupling factor is given by $I^{(0)}_1= \sqrt{N}|s^{(0)}|$ with $s^{(0)} = \sum_{n \in \mathcal{S}^{(0)}} c_n$ denoting the aggregate correlation.

\textbf{Step 2:} Next, the antennas are sequentially removed based on a principle of quasi-in-phase antenna deactivation. Supposing that $\ell$ antennas have been deactivated, the resulting interference factor is $I^{(\ell)}_1=\frac{N}{\sqrt{N - \ell}} \!|s^{(\ell)}|$ with $s^{(\ell)}=\sum_{n \in \mathcal{S}^{(\ell)}} c_n$.
The next antenna to be deactivated, i.e., $n^{(\ell+1)}_k$, is to maximize the interference reduction $I^{(\ell+1)}_{\Delta}$, which is defined as
		\begin{align}
			I^{(\ell+1)}_{\Delta} &\triangleq \! I^{(\ell)}_1 - I^{(\ell+1)}_1 \nonumber\\
			&=\!\frac{N}{\sqrt{N - \ell}} \!|s^{(\ell)}| - \frac{N}{\sqrt{N - \ell - 1}} \!|s^{(\ell)} - c_n|.
		\end{align}
   Since $I^{(\ell)}_1$ is determined by the previous step, maximizing $I^{(\ell+1)}_{\Delta}$ is equivalent to minimizing the residual correlation magnitude $|s^{(\ell)} - c_n|$, thereby leading to the following greedy deactivation principle: 
   
   \vspace{2pt}
   \noindent\vbox{\hrule\hbox{\vrule\kern0.5pt\vbox{\kern5pt
   			\begin{minipage}{\dimexpr\linewidth-12pt\relax} 
   				\centering
   				% 删除了 \small，现在字体大小与正文一致
   				\textbf{Greedy antenna selection for the two-user case:}
   				{\small
   					\begin{equation}
   						\label{eq:nf_r_samples}
   						\begin{aligned}
   						&n^{(\ell+1)} = \arg\min_{n \in \mathcal{S}^{(\ell)}} |\angle s^{(\ell)} - \angle c_n|,\;
   						\ell \in \{0,1,\cdots,N-1\}.
   						\end{aligned}
   				\end{equation}}
   			\end{minipage}
   			\kern0.5pt}\kern-2pt\vrule}\hrule}

\textbf{Step 3:} Update the activated antenna set and the resulting correlation as $\mathcal{S}^{(\ell+1)} = \mathcal{S}^{(\ell)} \setminus \{n^{(\ell+1)}\}$ and $s^{(\ell+1)} = s^{(\ell)} - c_{n^{(\ell+1)}}$, respectively. This process repeats until the number of remaining activated antennas reaches the target number $M_1$.

\begin{lemma} \label{lemma_I_dff}\rm
Supposing that $\ell$ antennas have been deactivated and $\ell \ll N$, the reduction of interference coupling factor during the $(\ell+1)$-th deactivation can be approximated as
	\begin{align}
		I^{(\ell+1)}_{\Delta} \approx \frac{\cos(\theta_{\Delta}^{(\ell+1)})}{\sqrt{N-\ell}},\label{Interference_reduction}
	\end{align}
where $\theta_{\Delta}^{(\ell+1)} = \angle s^{(\ell)} - \angle c_{	n^{(\ell+1)}}$. %and $n^{(\ell+1)}$ is the index of the deactivated antenna.
\end{lemma}
\begin{proof}
	Please refer to Appendix \ref{I_diff}. 
\end{proof}

\textbf{Lemma~\ref{lemma_I_dff}} reveals that the interference reduction at the $(\ell+1)$-th deactivation step depends on the phase difference $\theta_{\Delta}^{(\ell+1)}$ and the accumulated number of deactivated antennas $\ell$. However, it is challenging to explicitly characterize the relationship between the interference coupling factor and the number of deactivated antennas. To tackle this difficulty, we next numerically show the interference coupling factor obtained from the proposed low-complexity scheme, based on which, an efficient \emph{data regression} method is employed to characterize this relationship. The main procedures are summarized as follows.

\subsubsection{Model selection}
Based on the proposed antenna-selection scheme, we first plot the curves of $I_1^{(\ell)} $ vs. the number of deactivated antennas $\ell$ under various near-field user locations as shown in Figs.~\ref{Interference_rate_plot}(a) and 3(b), where the far-field user is located at (0~rad, 150~m) under the polar coordinate ($\theta,r$). Several important observations are made as follows.

(i) $0 \leq I_1^{(\ell)} \leq I_1^{(0)}$.

(ii) $I_1^{(\ell)}$ first decreases approximately linearly with $\ell$, and then reaches an interference floor with minor residual perturbations, when the interference level is sufficiently small.

(iii) $I_1^{(0)}$ serves as the $y$-axis intercept.

The above observations motivate us to approximate $I_1^{(\ell)}$ as a linear function with the number of deactivated antennas $\ell$
	\begin{align}
	I_1^{(\ell)} \approx \bar{I}_1^{(\ell)}  \triangleq I_1^{(0)} - \alpha_1\cdot \ell, \label{reduction_function_fitNF}
\end{align}
where $\alpha_1>0$ is a constant affected by the size of XL-array.

\subsubsection{Model evaluation}
\begin{figure}[t]
	\centering
	\begin{subfigure}[b]{0.49\linewidth}
		\includegraphics[width=1.105\linewidth]{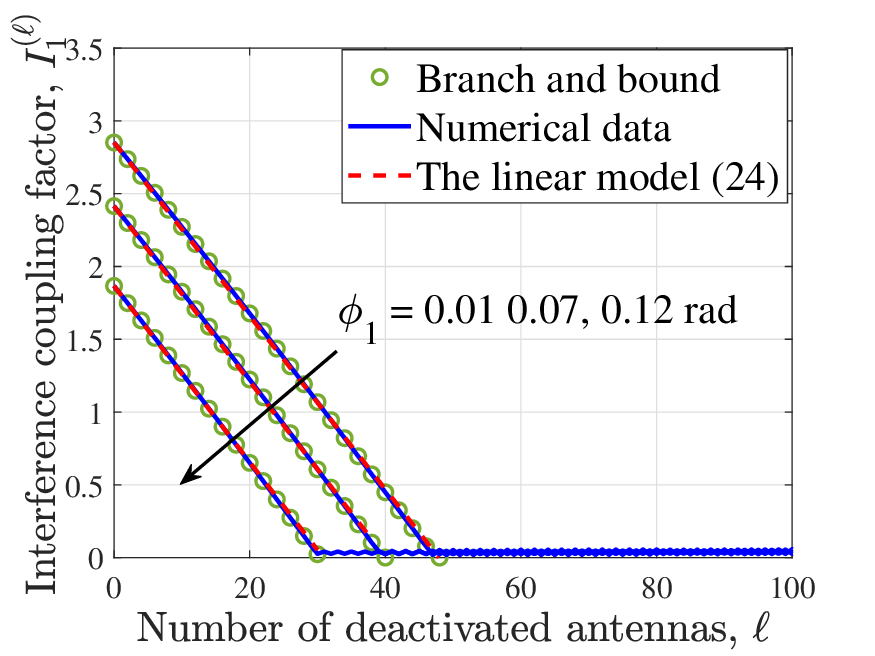}
		\caption{Antenna selection for near-field user under various angles $\phi_1$.}
		\label{Interference_vs_angle}
	\end{subfigure}
	\begin{subfigure}[b]{0.49\linewidth}
		\includegraphics[width=1.105\linewidth]{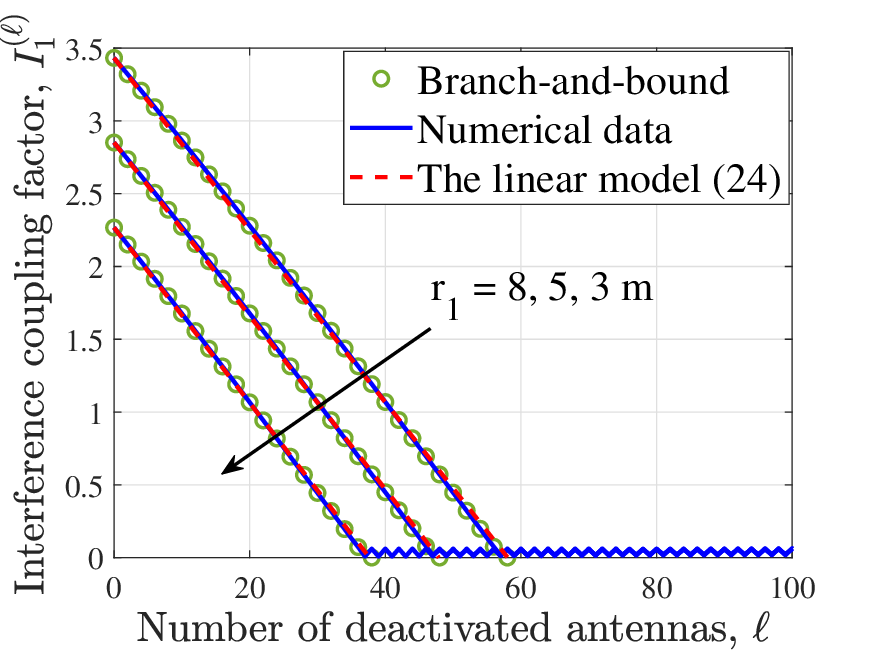}
		\caption{Antenna selection for near-field user under various distances $r_1$.}
		\label{Interference_vs_distance}
	\end{subfigure}
	\begin{subfigure}[b]{0.49\linewidth}
		\includegraphics[width=1.105\linewidth]{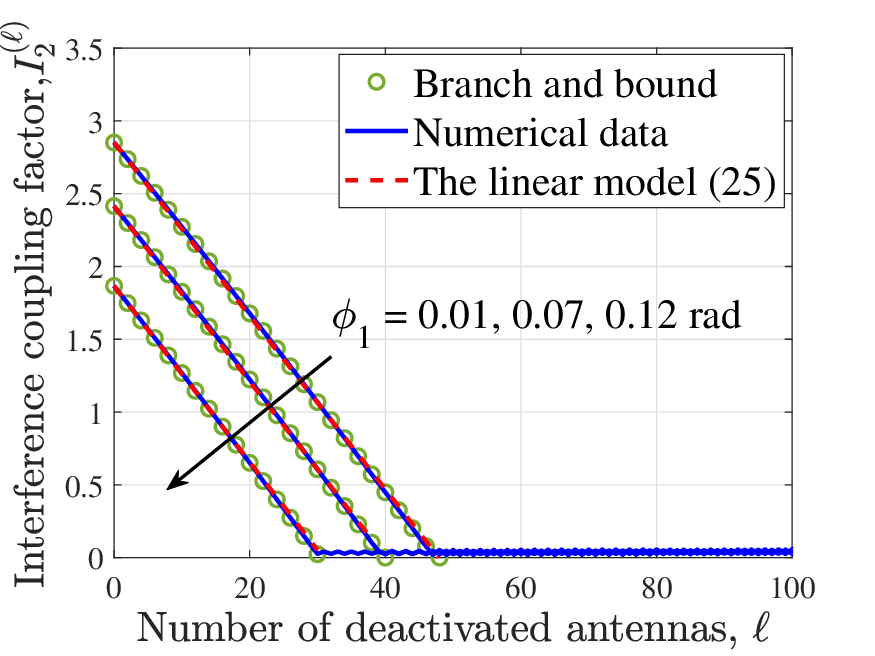}
		\caption{Antenna selection for far-field user under various angles $\phi_1$.}
		\label{Interference_analysis}
	\end{subfigure}
	\begin{subfigure}[b]{0.49\linewidth}
		\includegraphics[width=1.105\linewidth]{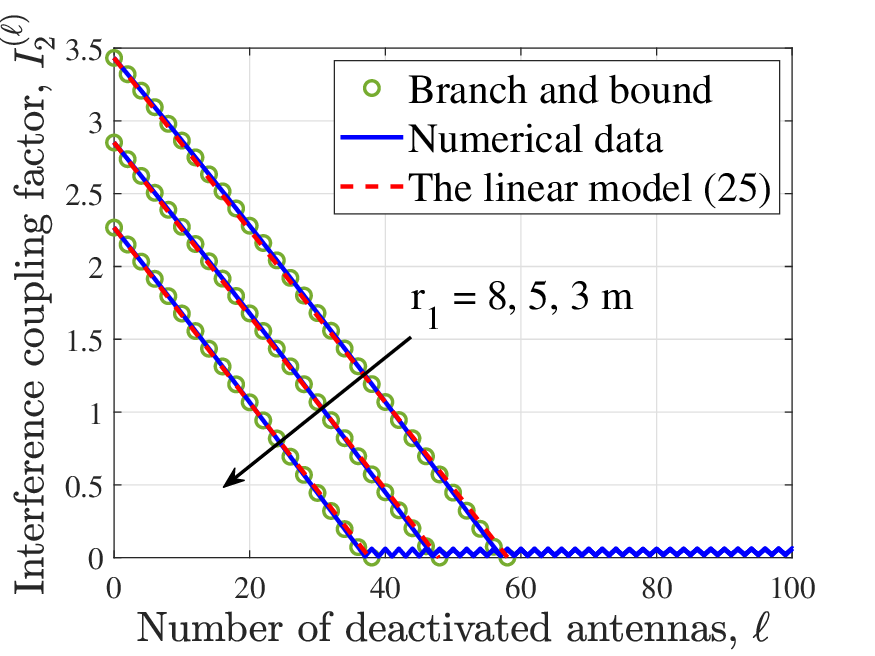}
		\caption{Antenna selection for far-field user under various distances $r_1$.}
		\label{Interference_closedform}
	\end{subfigure}
	\caption{Interference coupling factor vs. number of deactivated antennas.}
	\label{Interference_rate_plot}
\end{figure}

Next, to evaluate the accuracy of the linear model, the model parameter $\alpha_1$ is selected to closely match the numerical data based on the criterion of minimum mean square error. As shown in Figs.~\ref{Interference_rate_plot}(a) and 3(b), the proposed linear model in \eqref{reduction_function_fitNF} matches the numerical data very well. Regarding the antenna selection for the far-field user shown in Figs.~\ref{Interference_rate_plot}(c) and 3(d), a similar linear model can be adopted to characterize its interference level imposed on the near-field user, given by 
\begin{align}
	I_2^{(\ell)} \approx \bar{I}_2^{(\ell)} \triangleq I_2^{(0)} - \alpha_2 \cdot \ell.\label{reduction_function_fitFF}
\end{align} 

\begin{remark} \rm \label{remark_linear_decay}
	Based on \textbf{Lemma~\ref{lemma_I_dff}} and Fig.~\ref{Interference_rate_plot}, the properties of interference coupling factor are summarized as follows.
	\begin{itemize}
		\item For XL-array systems, the proposed greedy scheme progressively deactivates one antenna whose phase is best aligned with the aggregate correlation of the activated antenna set. As such, $\cos(\theta_{\Delta}^{(\ell+1)}) \approx 1$ and
		$1/\sqrt{N-\ell} \approx 1/\sqrt{N}$ hold for $\ell \ll N$, leading to a nearly constant interference reduction per step, i.e., $I^{(\ell+1)}_{\Delta} = \cos(\theta_{\Delta}^{(\ell+1)})/\sqrt{N-\ell} \approx 1/\sqrt{N}$. This leads to the linear decay of
		the interference coupling factor, whose slope is primarily determined by the array size $N$.
		\item When the initial near-far channel correlation (e.g., due to the smaller angular difference or larger near-field distance) is stronger, a larger number of antennas should be deactivated to achieve the same target interference level (see Fig.~\ref{Interference_rate_plot}).
	\end{itemize}
\end{remark}

\begin{remark} \rm (Performance optimality and computational complexity)
	Note that the antenna-selection problem can also be solved by using the MOSEK solver with a branch-and-bound framework to provide a globally optimal benchmark, as illustrated in Fig. \ref{Interference_rate_plot}(a). However, the computational complexity of branch-and-bound grows exponentially with the array size, i.e., $\mathcal{O}(2^N)$ in the worst case, which becomes computationally prohibitive for XL-arrays.
	Notably, the numerical results obtained by the proposed antenna-selection scheme closely match the MOSEK-based performance.
	This verifies the effectiveness and near-optimal performance of our proposed scheme, which enjoys a significantly lower computational complexity of $\mathcal{O}(N^2)$.
\end{remark}

\vspace{-15pt}
\subsection{Outer Problem: Optimization for Deactivated Antennas}
In this subsection, we solve the outer problem \textbf{(P2.2)} to optimize the number of deactivated antennas with its antenna selection method following the proposed low-complexity design in Section IV-A. Based on \eqref{reduction_function_fitNF} and \eqref{reduction_function_fitFF}, after deactivating $\ell_1=N-M_1$ and $\ell_2=N-M_2$ antennas for near-field and far-field users, respectively, their achievable rates in \eqref{AchinforRate_selec_uN} and \eqref{AchinforRate_selec_uF} can be approximated as follows 
	\begin{align}
		&R^{\rm AS}_1\!\approx\!\bar{R}^{\rm AS}_1\!\triangleq
		\!\log _2\!\!\left(\!\!1\!\!+\!\!\frac{P_1|h_1|^2(N-\ell_1)}{P_2|h_1|^2(I_2^{(0)} - \alpha_2 \ell_2)^2\!+\!\sigma^2}\!\!\right)\!\!,  
		\label{AchinforRate_approx_uN}\\
		&R^{\rm AS}_2 \!\approx\!\bar{R}^{\rm AS}_2\!\triangleq
		\!\log _2\!\!\left(\!\!1\!\!+\!\!\frac{P_2|h_2|^2(N-\ell_2)}{P_1|h_2|^2(I_1^{(0)} - \alpha_1 \ell_1)^2\!+\!\sigma^2}\!\!\right)\!\!,   \label{AchinforRate_approx_uF}
	\end{align}
and thus Problem \textbf{(P2.2)} can be reformulated as
\begin{align}
	\textbf{(P3)} \;\;
	\max_{\substack{\{\ell_k\}}}  \quad&\bar{R}^{\rm AS}_1+\bar{R}^{{\rm AS}}_2\nonumber\\
	\text{s.t.}\quad\;&\ell_k \in \{0,1,\cdots, N-1\},\forall k\in \{1,2\}. \label{InCon}
\end{align}

It is observed from \eqref{AchinforRate_approx_uN} and \eqref{AchinforRate_approx_uF} that increasing $\ell_k$ gradually reduces the inter-user interference coupling factor, i.e., $I_k^{(0)}\!-\!\alpha_k \ell_k$, but also inevitably reduces the effective array gain $N\!-\ell_k$. Therefore, there is a tradeoff in antenna selection between interference suppression and array-gain enhancement for the sum-rate maximization. For XL-array systems where the desired-signal power dominates noise and interference, the achievable rates can be further
approximated as
\begin{align}
	\bar{R}^{\rm AS}_1
	&\approx \tilde{R}^{\rm AS}_1=\log_2\left(\frac{P_1|h_1|^2(N-\ell_1)}
	{P_2|h_1|^2(I_2^{(0)}-\alpha_2\ell_2)^2+\sigma^2}\right),\label{AchinforRate_approx_uN1}
	\\
	\bar{R}^{\rm AS}_2
	&\approx \tilde{R}^{\rm AS}_2	=\log_2\left(\frac{P_2|h_2|^2(N-\ell_2)}
	{P_1|h_2|^2(I_1^{(0)}-\alpha_1\ell_1)^2+\sigma^2}\right).\label{AchinforRate_approx_uF1}
\end{align}
Furthermore, by relaxing the integer constraint in \eqref{InCon},
Problem \textbf{(P3)} can be approximated as
\begin{align}
	\textbf{(P3.1)}\;\max_{\{\ell_k\}}\;\tilde{R}^{\rm AS}_1\!+\!\tilde{R}^{\rm AS}_2 \;\;
	\text{s.t.}\;0\!\le\!\ell_k\!\le\! N\!-\!1,\forall k\!\in\!\{\!1,2\}\!.
\end{align}
 The optimal solution to the relaxed problem~\textbf{(P3.1)} can be obtained as follows.
\begin{lemma}\rm \label{Lemma_optimal_number}
	The optimal solution to \textbf{(P3.1)} is given by
	\begin{align}
		\ell_1 &= N-\sqrt{\left(N-\frac{I_1^{(0)}}{\alpha_1}\right)^2
			+\frac{\sigma^2}{\alpha_1^2|h_2|^2P_1}}, \label{Num_Opt_NF}\\
		\ell_2 &= N-\sqrt{\left(N-\frac{I_2^{(0)}}{\alpha_2}\right)^2
			+\frac{\sigma^2}{\alpha_2^2|h_1|^2P_2}}.\label{Num_Opt_FF}
	\end{align}
	Based on integer rounding, a suboptimal solution for the number of deactivated antennas is obtained as
	\begin{align}
		\ell^{\star}_1 = \lfloor \ell_1 \rfloor, \qquad
		\ell^{\star}_2 = \lfloor \ell_2 \rfloor.
	\end{align}
\end{lemma}
\begin{proof}
	Please refer to Appendix \ref{proof_optimal_number}.
\end{proof}

\begin{figure}[t]
	\centering
	\begin{subfigure}[b]{0.49\linewidth}
		\includegraphics[width=1.11\linewidth]{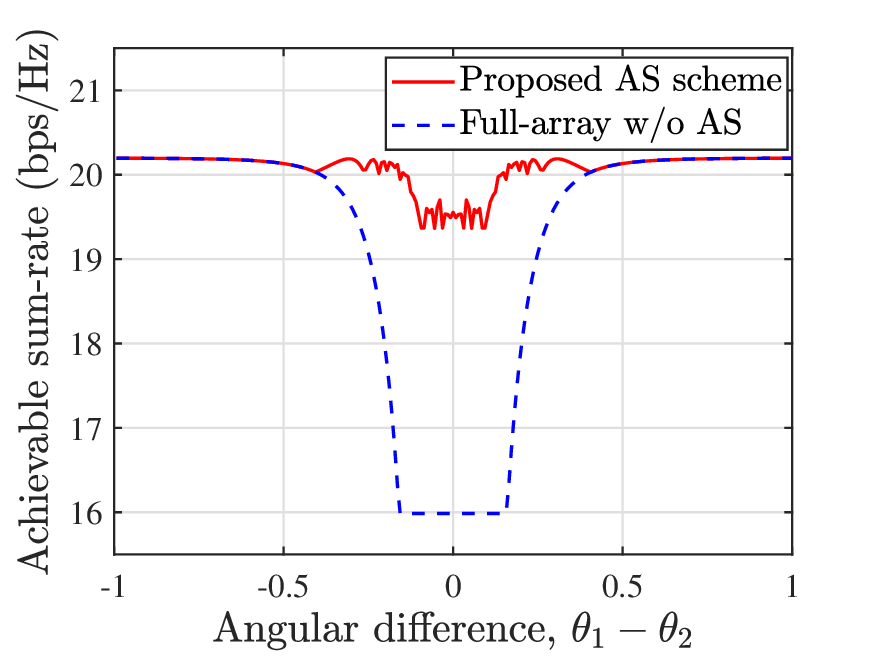}
		\caption{Achievable sum-rate.}
	\end{subfigure}
	\begin{subfigure}[b]{0.49\linewidth}
		\includegraphics[width=1.11\linewidth]{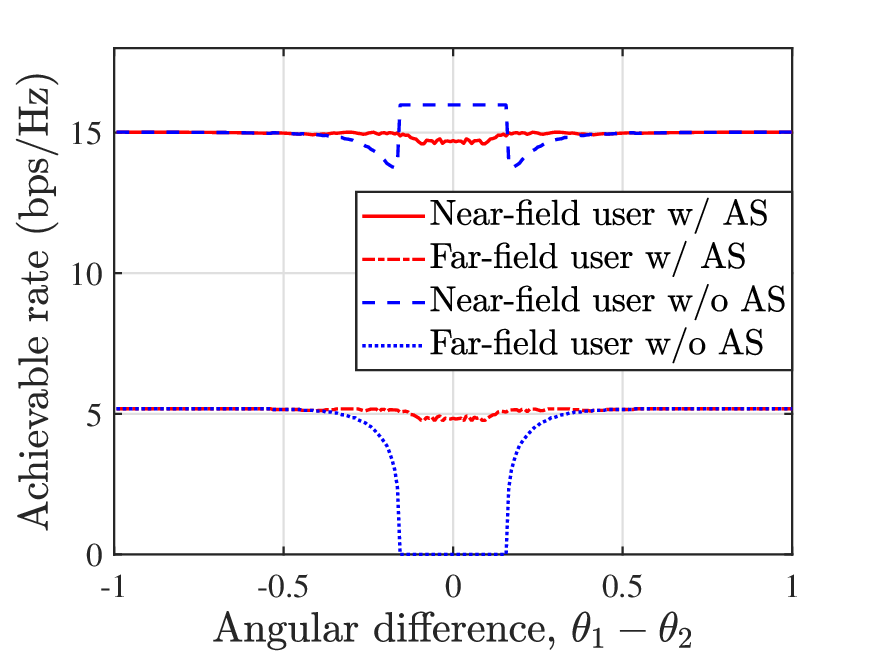}
		\caption{Individual user rates.}
	\end{subfigure}
	\caption{Illustration of the performance obtained by the proposed antenna deactivation scheme with optimized power allocation.}
	\label{gain_with_angle}
\end{figure}

\begin{example} \rm
	To demonstrate the performance of the proposed antenna-deactivation scheme, we consider a two-user mixed-field scenario with one near-field user located at $(0~\text{rad}, 5~\text{m})$ and one far-field user located at $(0.05~\text{rad}, 150~\text{m})$. Specifically, power allocation and antenna selection are jointly optimized via alternating optimization. For given antenna-selection matrices, the transmit power is optimized via a one-dimensional search, while the antenna-selection matrices are updated using the proposed low-complexity design.
	We plot in Fig.~\ref{gain_with_angle}(a) the achievable sum-rate versus the angular difference between the two users. One can observe that the proposed antenna-selection scheme significantly outperforms the full-array baseline when the far-field user is located within an angular region close to the near-field user (i.e., around $0$~rad), owing to its interference-suppression capability. 
	To provide further insight, Fig.~\ref{gain_with_angle}(b) shows the achievable rates of the near-field and far-field users, respectively. One can observe that when the angular difference between the users becomes small, the achievable rate of the far-field user in the full-array scheme rapidly degrades and approaches zero due to strong mixed-field interference. In contrast, our proposed scheme maintains a stable and substantial rate for the far-field user. 
\end{example}

\vspace{-5pt}
\section{General Case: Proposed Algorithm for Solving Problem (P1)}
In this section, we consider the general multi-user case. Specifically, we first develop a PDD-based double-layer optimization algorithm to efficiently solve Problem \textbf{(P1)}. Then, a low-complexity yet efficient algorithm is proposed to obtain its high-quality solution.

\vspace{-5pt}
\subsection{Proposed PDD-based Algorithm}

\subsubsection{Problem reformulation for Problem \textbf{(P1)}}

For the multi-user case, we first use $\mathbf{v}_k$ to collect the diagonal entries of $\mathbf{V}_k$, i.e.,  $[\mathbf{v}_k]_n= [\mathbf{V}_k]_{n,n}$. As such, we have $\mathbf{h}_k^H \mathbf{V}_i \mathbf{w}_i=\mathbf{t}_{k,i}^H\mathbf{v}_i$ with $\mathbf{t}_{k,i}^H = \mathbf{h}_k^H {\rm diag}(\mathbf{w}_i), \forall k,i \in \mathcal{K}$, leading to the following equivalent form for the achievable rate in \eqref{RateUnify}:
\begin{align}
	\hat{R}^{{\rm AS}}_{k} \!= \!\log_2\!\!\left(\!1 \!+\! \frac{\frac{P_k }{M_k} \left| \mathbf{t}_{k,k}^H\mathbf{v}_k  \right|^2}{\sum^K_{i=1,i \neq k}\!\! \frac{P_i }{M_i} \!\left|\mathbf{t}_{k,i}^H  \mathbf{v}_i \right|^2\! +\! \sigma^2} \!\right)\!\!,\forall k \in \mathcal{K}.
\end{align}
To facilitate the optimization, we explicitly treat $\{M_k\}$ as an optimization variable by decoupling it from $\mathbf{v}_k$ through an equality constraint. Then, Problem \textbf{(P1)} becomes
\begin{align}
	\textbf{(P4)}\;\; \max_{\{P_k,\mathbf{v}_k,M_k\}} \quad &\sum^K_{k=1} \hat{R}^{{\rm AS}}_{k} \nonumber\\
	\text{s.t.}\;\;\;\;\;
	&[\mathbf{v}_k]_n  \in \{0,1\}, \forall n \in \mathcal{N}, \forall k \in \mathcal{K}, \label{P3AntennaSelection}\\
	&\sum^N_{n=1}[\mathbf{v}_k]_n  = M_k, \forall k \in \mathcal{K},\label{P3SumNumAnten}\\
	&M_k \! \in \! \{0,1,2,\cdots,N\}, \forall k \in \mathcal{K}, \label{LisanAntennaNum}\\
	&\eqref{TotPower}. \nonumber
\end{align}

To solve this problem, we introduce auxiliary variables $\{\tilde{\mathbf{v}}_k,\forall k\in \mathcal{K}\}$ and enforce $[\tilde{\mathbf{v}}_k]_n= [\mathbf{v}_k]_n ,\forall k,n$. As such, the binary constraints in \eqref{P2AntennaSelection} can be equivalently transformed into the following set of constraints:
\begin{align}
	&[\mathbf{v}_k]_n\left(1-[\tilde{\mathbf{v}}_k]_n\right)=0, \forall n \in \mathcal{N}, \forall k \in \mathcal{K}, \label{IntroVhatBinary}\\
	&[\mathbf{v}_k]_n = [\tilde{\mathbf{v}}_k]_n, \forall n \in \mathcal{N}, \forall k \in \mathcal{K}, \label{IntroVhatEqual}\\
	&0\leq[\mathbf{v}_k]_n\leq 1,\forall n \in \mathcal{N}, \forall k \in \mathcal{K}.\label{IntroVhatRange}
\end{align}

Besides, the discrete constraint \eqref{LisanAntennaNum} is relaxed to a continuous counterpart as
\begin{align}
	0 \leq M_k \leq N, \;\forall k \in \mathcal{K}. \label{slackNk}
\end{align}
Furthermore, by incorporating the equality constraints \eqref{P3SumNumAnten}, \eqref{IntroVhatBinary} and \eqref{IntroVhatEqual} into the objective function, the augmented
Lagrangian form associated with Problem \textbf{(P4)} is given by~\cite{Shi_Qingjiang2020}
\begin{align}
	\textbf{(P5)}\;
	\max_{\mathbb{V}}\quad
	&\sum_{k=1}^{K} R_k -L_1(\mathbf{v}_k, M_k, \boldsymbol{\mu}) \nonumber\\
	&\quad\quad\quad\quad\quad-L_2(\mathbf{v}_k, \tilde{\mathbf{v}}_k, \boldsymbol{\delta}) - L_3(\mathbf{v}_k, \tilde{\mathbf{v}}_k, \boldsymbol{\lambda})
	\nonumber\\
	\text{s.t.}\quad&\eqref{TotPower},  \eqref{IntroVhatRange}, \eqref{slackNk}, \nonumber
\end{align}
where $\mathbb{V}=\{P_k, \mathbf{v}_k, \tilde{\mathbf{v}}_k,  M_k\}$ denotes the set of optimization variables. The penalty terms $L_1$, $L_2$ and $L_3$ are respectively given by
\begin{align}
		&L_1(\mathbf{v}_k, M_k, \boldsymbol{\mu})\!=\!\frac{1}{2 \rho} \sum_{k=1}^K\Bigl(\sum_{n=1}^N [\mathbf{v}_k]_n-\!M_k\!+\!\rho \mu_k\Bigr)^2\!\!, \\
	&L_2(\mathbf{v}_k, \tilde{\mathbf{v}}_k, \boldsymbol{\delta})\!=\!\frac{1}{2 \rho} \sum_{k=1}^K \sum_{n=1}^N\left([\mathbf{v}_k]_n-[\tilde{\mathbf{v}}_k]_n\!+\!\rho \delta_{k,n}\right)^2\!\!,\\
	&L_3(\mathbf{v}_k, \tilde{\mathbf{v}}_k, \boldsymbol{\lambda})\!=\!\frac{1}{2 \rho} \sum_{k=1}^K\sum_{n=1}^N\!\left([\mathbf{v}_k]_n\left(1\!-\![\tilde{\mathbf{v}}_k]_n\right)\!+\!\rho \lambda_{k,n}\right)^2\!\!,
\end{align} 
with $\rho$ being the penalty parameter, and $\boldsymbol{\mu} \triangleq \{\mu_k, \forall k\in\mathcal{K}\}$, $\boldsymbol{\delta}\triangleq\{\delta_{k,n},\forall k\in\mathcal{K}, \forall n\in \mathcal{N}\}$, and $\boldsymbol{\lambda}\triangleq\{\lambda_{k,n}, \forall k\in\mathcal{K}, \forall n\in \mathcal{N}\}$ being the dual variables. 

However, \textbf{(P5)} is still a non-convex optimization problem due to the non-convex objective function and the intricate coupling among optimization variables. To address this issue, we propose a two-layer optimization framework \cite{Shi_Qingjiang2020,Guo_Rongbin2018}, where  the BCD method is employed to iteratively optimize  variables in $\mathbb{V}$ in the inner layer,  while the dual variables and penalty parameter are  updated in the outer layer.

\subsubsection{Inner-layer optimization of $\mathbb{V}$}To solve Problem \textbf{(P5)}, the variables in $\mathbb{V}$ are optimized alternately through the following four steps.
%\vspace{10pt}

\underline{\textbf{Optimization of $\{\mathbf{v}_k\}$:}} Given any feasible $\{\tilde{\mathbf{v}}_k\}$, $\{M_k\}$, and $\{P_k\}$, Problem (P5) reduces to the following problem for optimizing $\{\mathbf{v}_k\}$:
\begin{align}
	\textbf{(P6)}\;
	\max_{\left\{\mathbf{v}_k\right\}}  \quad
	&\sum_{k=1}^{K} R_k -L_1(\mathbf{v}_k, M_k, \boldsymbol{\mu}) \nonumber\\
	&\quad\quad\quad\quad\quad-L_2(\mathbf{v}_k, \tilde{\mathbf{v}}_k, \boldsymbol{\delta}) - L_3(\mathbf{v}_k, \tilde{\mathbf{v}}_k, \boldsymbol{\lambda})
	\nonumber\\
	\text { s.t. } \quad 
	& \eqref{IntroVhatRange}. \nonumber
\end{align}
Problem \textbf{(P6)} is a non-convex optimization problem due to the complex objective function. To tackle this difficulty, a set of auxiliary variables $\{\eta_k,\forall k \in \mathcal{K}\}$ are introduced to replace $R_k$ in the objective function, leading to the following constraints:
\begin{align}
	R_k \geq \eta_k, \forall k \in \mathcal{K}. \label{P2SlackForRate}
\end{align}
Furthermore, by introducing auxiliary variables $a_k\!=\!\sum^K_{i=1} \frac{P_i}{M_i} \big| \mathbf{t}_{k,i}^H  \mathbf{v}_i \big|^2$ and $b_k\!=\!\sum_{i=1, i \neq k}^{K} \!\!\frac{P_i}{M_i} \!\big| \mathbf{t}_{k,i}^H \mathbf{v}_i \big|^2$,
constraints \eqref{P2SlackForRate} can be equivalently re-expressed as the following constraints:
\begin{align}
	&~~~\log_2\left( a_k + \sigma^2\right)-\log_2\left(b_k + \sigma^2 \right) \geq \eta_k,\forall k \in \mathcal{K}, \label{LogDiffSlack}\\
	&~~\sum^K_{i=1} \frac{P_i  }{M_i} \left| \mathbf{t}_{k,i}^H  \mathbf{v}_i \right|^2 \geq a_k, \forall k \in \mathcal{K},\label{FracSlackak}\\
	&\sum^K_{i=1, i \neq k} \frac{P_i }{M_i} \left| \mathbf{t}_{k,i}^H  \mathbf{v}_i \right|^2 \leq b_k, \forall k \in \mathcal{K}. \label{FracSlackbk}
\end{align}

\begin{lemma} \label{FirstTylorlog} \rm
	For the constraint \eqref{LogDiffSlack}, the function $\zeta_k\triangleq\log_2\left(b_k + \sigma^2 \right)$ is a concave function of $b_k$. For any local point $\hat{b}_k$, $\zeta_k$ in \eqref{LogDiffSlack} can be upper-bounded by
	\begin{align}
		\zeta_k \leq \log_2\left(\hat{b}_k+\sigma^2\right)+\frac{\log_2(e)}{\hat{b}_k+\sigma^2}\left(b_k-\hat{b}_k\right) \triangleq  \zeta_k^{\rm ub}. \label{logTylor}
	\end{align}
\end{lemma}

\begin{lemma} \label{FirstTylorNorm} \rm
	For the constraint \eqref{FracSlackak}, the function $\psi_{k,i}\triangleq\big| \mathbf{t}_{k,i}^H\mathbf{v}_i \big|^2 $ is a convex function of $\mathbf{v}_i$. For any local point $\hat{\mathbf{v}}_i$, $\psi_{k,i}$ in \eqref{FracSlackak} can be lower-bounded by
	\begin{align}
		\psi_{k,i} \geq 2\mathcal{R}\{\hat{\mathbf{v}}^H_i\mathbf{t}_{k,i}\mathbf{t}_{k,i}^H\mathbf{v}_i\}- \left|\mathbf{t}_{k,i}^H\hat{\mathbf{v}}_i \right|^2 \triangleq  \psi_{k,i} ^{\rm lb}.
	\end{align}
\end{lemma}

Based on the above, Problem \textbf{(P6)} can be approximated by the following problem:
\begin{align}
	\textbf{(P6.1)}\;
	\max_{\substack{\{\mathbf{v}_k,\eta_k\\ a_k, b_k\}}}\quad
	&\sum_{k=1}^{K} \eta_k -L_1(\mathbf{v}_k, M_k, \boldsymbol{\mu}) \nonumber\\
	&\quad\quad\quad\quad\quad-L_2(\mathbf{v}_k, \tilde{\mathbf{v}}_k, \boldsymbol{\delta}) - L_3(\mathbf{v}_k, \tilde{\mathbf{v}}_k, \boldsymbol{\lambda})
	\nonumber\\
	\text{ s.t.}\quad
	&\log_2 \left( a_k\! +\! \sigma^2 \right) -\zeta_k^{\rm ub} \geq \eta_k,\forall k \in \mathcal{K}, \label{P51LogDiffSlack}\\
	&\sum^K_{i=1} \frac{P_i}{M_i} \psi_{k,i} ^{\rm lb} \geq a_k, \forall k \in \mathcal{K},\label{P51FracSlackak}\\
	&\eqref{IntroVhatRange},\eqref{FracSlackbk}. \nonumber
\end{align}
This is a convex optimization problem, which can be solved by using the CVX solvers. %~\cite{cvx}

\underline{\textbf{Optimization of $\{\tilde{\mathbf{v}}_k\}$:}}
Given any feasible $\{\mathbf{v}_k\}$, $\{M_k\}$, and $\{P_k\}$, Problem \textbf{(P5)} reduces to the following problem:% for optimizing $\{\tilde{\mathbf{v}}_k\}$
\begin{align}
	\textbf{(P7)}\;\;
	\min_{\left\{\tilde{\mathbf{v}}_k\right\}}&\quad L_2(\mathbf{v}_k, \tilde{\mathbf{v}}_k, \boldsymbol{\delta}) + L_3(\mathbf{v}_k, \tilde{\mathbf{v}}_k, \boldsymbol{\lambda}). \nonumber
\end{align}
which is also convex and can be efficiently solved by using the CVX solvers.  

 \underline{\textbf{Optimization of $\{M_k\}$:}}
Given any feasible $\{\mathbf{v}_k\}$, $\{\tilde{\mathbf{v}}_k\}$, and $\{P_k\}$ and
following a similar procedure as those in \eqref{P2SlackForRate}--\eqref{logTylor} for optimizing  ${\mathbf{v}_k}$, Problem \textbf{(P5)} reduces to
\begin{align}
	\textbf{(P8)}\;\;
	\max_{\left\{M_k,\eta_k, a_k, b_k\right\}} &\quad
	\sum_{k=1}^{K}\eta_k - L_1(\mathbf{v}_k, M_k, \boldsymbol{\mu})  \nonumber\\
	\text { s.t. } 
	&\quad \eqref{slackNk},\eqref{LogDiffSlack}-\eqref{FracSlackbk}. \nonumber
\end{align}
Problem \textbf{(P8)} is still non-convex  due to the constraints \eqref{LogDiffSlack} and \eqref{FracSlackak}.  Note that constraint \eqref{LogDiffSlack} can be approximated by the same convex form as in \eqref{P51LogDiffSlack} via \textbf{Lemma \ref{FirstTylorlog}}. For constraint \eqref{FracSlackak}, we employ the SCA technique to approximate it as follows.
\begin{lemma} \label{FirstTylorN} \rm
	For the constraint \eqref{FracSlackak}, the function $\xi_{i}\triangleq\frac{1}{M_i}$ is a convex function of $M_i$. For any local point $\hat{M}_i$, $\xi_{i}$ in \eqref{FracSlackak} can be lower-bounded by
	\begin{align}
		\xi_{i} \geq \frac{1}{\hat{M}_i}-\frac{1}{\hat{M}_i^2}\left(M_i-\hat{M}_i\right)  \triangleq  \xi_{i} ^{\rm lb}. 
	\end{align}
\end{lemma}

Based on \textbf{Lemmas \ref{FirstTylorlog}}  and \textbf{\ref{FirstTylorN}}, by substituting  $\zeta_k$ in \eqref{LogDiffSlack} and $\xi_{i}$ in \eqref{FracSlackak} with their corresponding bounds, Problem \textbf{(P8)} can be transformed into the following approximate form:
\begin{align}
	\textbf{(P8.1)}\;
		\max_{\substack{\{M_k,\eta_k\\ a_k, b_k\}}}
    & \sum_{k=1}^{K}\eta_k  \!-\! L_1(\mathbf{v}_k, M_k, \boldsymbol{\mu}) \nonumber\\
	\text{ s.t.}\;\;\;\;\;\;
	&\sum^K_{i=1} \! P_i N \! \left| \mathbf{t}_{k,i}^H  \mathbf{v}_i \right|^2\!\xi_{i} ^{\rm lb} \geq a_k, \forall k \in \mathcal{K},\label{P72FracSlackak}\\
	& \eqref{slackNk},\eqref{FracSlackbk},\eqref{P51LogDiffSlack}. \nonumber
\end{align}
The CVX solver can be directly applied to efficiently solve this convex problem. 

\underline{\textbf{Optimization of $\{P_k\}$:}} Last, given any feasible $\{\mathbf{v}_k\}$, $\{\tilde{\mathbf{v}}_k\}$, and $\{M_k\}$, the problem for optimizing $\{P_k\}$ can be formulated as
\begin{align}
	\textbf{(P9)}\;\;
	\max_{\left\{P_k,\eta_k, a_k, b_k\right\}}~\sum_{k=1}^{K} \eta_k \quad
	\text { s.t. }~\eqref{TotPower}, \eqref{FracSlackak},\eqref{FracSlackbk},\eqref{P51LogDiffSlack}. \nonumber
\end{align}
Note that the procedure for optimizing $\{ P_{k}\}$ follows the same procedures as those in \eqref{P2SlackForRate}--\eqref{logTylor} for optimizing $\{ \mathbf{v}_{k}\}$, thus we omit it here for brevity.

Based on the above, a suboptimal solution to \textbf{(P5)} can be obtained by iteratively solving problems \textbf{(P6.1)},  \textbf{(P7)}, \textbf{(P8.1)}, and \textbf{(P9)} until convergence is achieved.

\subsubsection{Outer-layer update of dual variables and penalty parameter}
In the outer iteration, the penalty coefficient $\rho$ is updated by
\begin{align}
\rho^{(i_{\rm o}+1)}=c\rho^{(i_{\rm o})}, \label{update_rho}
\end{align}
where $c$ is a constant scaling factor and $i_{\rm o}$ denotes the outer iteration index. The dual variables $\{\lambda_{k,n}, \delta_{k,n}, \mu_k\}$ are updated as follows:
\begin{align}
	&\lambda^{(i_{\rm o}+1)}_{k,n} \!=\! \lambda^{(i_{\rm o})}_{k,n}\!+\!\frac{1}{\rho^{(i_{\rm o})}}\big([\mathbf{v}_k]_n (1-[\tilde{\mathbf{v}}_k]_n)\big),\forall n,k, \label{update_lambda}\\
	&\delta^{(i_{\rm o}+1)}_{k,n}\!=\!\delta^{(i_{\rm o})}_{k,n}\!+\!\frac{1}{\rho^{(i_{\rm o})}}\big([\mathbf{v}_k]_n-[\tilde{\mathbf{v}}_k]_n\big),\forall n, k,\label{update_delta}\\
	&\mu^{(i_{\rm o}+1)}_k \!=\! \mu^{(i_{\rm o})}_k\!+\!\frac{1}{\rho^{(i_{\rm o})}}\big(\sum_{n=1}^N[\mathbf{v}_k]_n-M_k\big), \forall k .\label{update_mu}
\end{align}  
We define the constraint violation indicator $g$ as
\begin{align}
	g = \max\{ &|[\mathbf{v}_k]_n\!\left(1-[\tilde{\mathbf{v}}_k]_n\right)|, |[\mathbf{v}_k]_n-[\tilde{\mathbf{v}}_k]_n|, \nonumber\\ &\qquad\; |\sum_{n=1}^N[\mathbf{v}_k]_n-M_k|, \forall n \in \mathcal{N}, \forall k \in \mathcal{K}\} \label{constraint_violation}.
\end{align}
Note that the algorithm terminates when $g$ is smaller than a given threshold $\varepsilon$~\cite{Shi_Qingjiang2020}, indicating that the equality constraints \eqref{P3SumNumAnten}, \eqref{IntroVhatBinary} and \eqref{IntroVhatEqual} are almost satisfied.

\begin{remark}\rm (Algorithm convergence and computational complexity)
	For the proposed PDD-based two-layer algorithm for solving Problem \textbf{(P1)}, its convergence is analyzed as below. First, in the outer layer, the penalty parameter and dual variables are updated based on \eqref{update_rho}--\eqref{update_mu}. As the penalty factor $\rho$ decreases, the equality constraints \eqref{P3SumNumAnten}, \eqref{IntroVhatBinary} and \eqref{IntroVhatEqual} are gradually enforced. In the inner layer, Problem \textbf{(P5)} is solved by alternately solving convex subproblems, whose objective is non-decreasing and upper-bounded, thus ensuring algorithm convergence. In terms of computational complexity, the burden of proposed algorithm is primarily dominated by the subproblems for antenna-selection vectors $\{\mathbf{v}_k, \tilde{\mathbf{v}}_k\}$, leading to an overall complexity that scales as $O(I_{\rm in}I_{\rm out}(KN)^{3.5})$, where $I_{\rm in}$ and $I_{\rm out}$ denote the number of iterations to achieve convergence in the inner-layer and outer-layer, respectively. 
\end{remark}

\vspace{-15pt}
\subsection{Proposed Low-complexity Algorithm}

To reduce the computational complexity of the antenna selection design, we further develop a low-complexity algorithm in this subsection for the multi-user case that first designs antenna selection based on the multi-user quasi-in-phase antenna deactivation, followed by the second phase of power allocation. Note that the proposed algorithm for the two-user case in Section IV can be treated as a special case of the proposed algorithm below.

\subsubsection{Interference-suppression-based antenna selection}

Similar to the two-user case in Section IV, we consider antenna selection design for each user $k$, for which the objective is to minimize its total interference imposed on the other $K-1$ users, i.e., minimize $	I_k=\sum^{K}_{i=1, i \neq k} \frac{1}{\sqrt{M_k}} |\mathbf{h}_i^H \mathbf{V}_k \mathbf{w}_k|$. The algorithm is composed of the following four procedures.

\textbf{Step 1:} For each user $k$, we first evaluate the contribution of the $n$-th antenna to the interference imposed on user $i$, defined as $z_{k,i,n} = [\mathbf{h}_i]_n^{*}[\mathbf{w}_k]_n$ for $i \in \mathcal{K} \setminus \{k\}$. For an initial set of activated antennas $\mathcal{S}^{(0)}_k=\{1,2,\ldots,N\}$, the interference coupling factor is given by
	\begin{align}
			I_k(\mathcal{S}^{(0)}_k)=\sum_{i=1,\, i \neq k}^{K}\frac{1}{\sqrt{N}}\Bigg| \sum_{n \in \mathcal{S}^{(0)}_k} z_{k,i,n}\Bigg|,\forall k \in \mathcal{K}.\label{eq:IkS}
	\end{align}

\textbf{Step 2:} Next, the antennas are progressively deactivated based on the quasi-in-phase antenna deactivation principle. Supposing that $\ell$ antennas have been removed, we let $\mathcal{S}^{(\ell)}_k$ denote the set of remaining activated antennas. The next antenna to be deactivated, i.e., $n^{(\ell+1)}_k$, is to minimize the residual interference $n^{(\ell+1)}_k = \arg\min_{n \in \mathcal{S}^{(\ell)}_k} I_k\big(\mathcal{S}^{(\ell)}_k \setminus \{n\}\big)$, after which the activated antenna set is updated as $\mathcal{S}^{(\ell+1)}_k = \mathcal{S}^{(\ell)}_k \setminus \{n^{(\ell+1)}_k\}$. 
This procedure repeats until there is only one antenna, yielding a candidate sequence $\{\mathcal{S}^{(0)}_k,\dots,\mathcal{S}^{(N-1)}_k\}$.

	\vspace{2pt}
	\noindent\vbox{\hrule\hbox{\vrule\kern0.5pt\vbox{\kern5pt
				\begin{minipage}{\dimexpr\linewidth-12pt\relax} 
					\centering
					% 删除了 \small，现在字体大小与正文一致
					\textbf{Greedy selection principle in multi-user scenarios:}
					{\small
						\begin{equation}
							\label{principle_multiuser}
							\begin{aligned}
							&n^{(\ell+1)}_k\!\! =\! \arg\!\!\min_{n\in\mathcal{S}^{(\ell)}_k}\!\!\! \Bigg(\!\sum^{K}_{i=1, i \neq k} \!\!\frac{1}{\sqrt{N-\ell}}\! \Bigg|\!\sum_{m \in \mathcal{S}^{(\ell)}_k} \!\! z_{k,i,m}\!-\!z_{k,i,n}\!\Bigg|\!\Bigg)\!,\\
							& \forall k \in \mathcal{K}, \ell \in \{0,1,\cdots,N-1\}.
							\end{aligned}
					\end{equation}}
				\end{minipage}
				\kern0.5pt}\kern-2pt\vrule}\hrule}

\textbf{Step 3:} Finally, the candidate subset that minimizes the total interference is selected as the final activated antenna set for user $k$, i.e., $\mathcal{S}^{(\ell^{\star})}_k$, with $\ell^{\star}=\arg\min_{\ell \in \{0,1,\ldots,N-1\}}I_k(\mathcal{S}^{(\ell)}_k)$.\footnote{This interference-minimization criterion is an efficient heuristic that decouples the multi-user antenna selection, significantly reducing complexity while maintaining high sum-rate performance, as shown in Section~VI.}

\begin{figure}[t]
	\centering
	\begin{subfigure}[b]{0.49\linewidth}
		\includegraphics[width=1.02\linewidth]{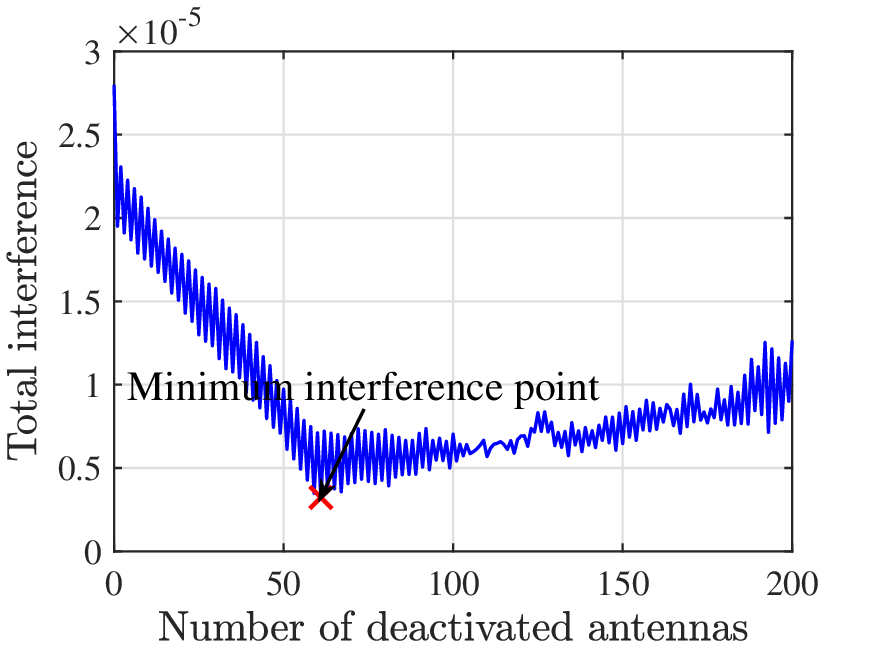}
		\caption{Representative near-field user~1.}
	\end{subfigure}
	\begin{subfigure}[b]{0.49\linewidth}
		\includegraphics[width=1.02\linewidth]{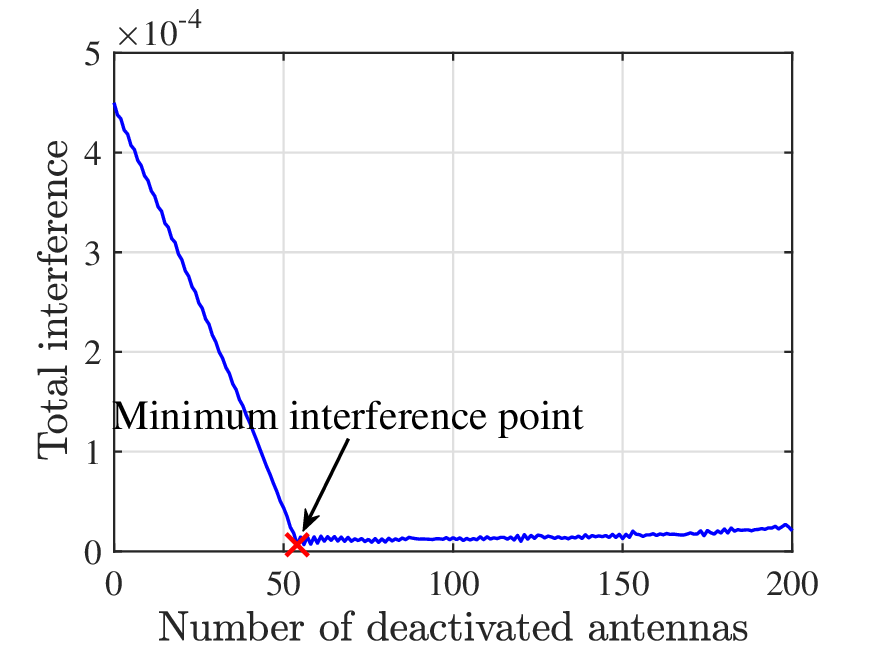}
		\caption{Representative far-field user~3.}
	\end{subfigure}
	\caption{Total interference during greedy deactivation in a five-user system: 2~near-field users at $(-0.30, 5.82\,\text{m})$ and $(0.17, 6.21\,\text{m})$, and 3~far-field users at $(-0.22, 150\,\text{m})$, $(0.09, 175\,\text{m})$, and $(0.25, 200\,\text{m})$.}
	\label{Interfer_with_DeNum_Multiuser}
\end{figure}

\begin{remark} \rm Under the proposed greedy deactivation principle, the total interference $I_k(\mathcal{S}_k^{(\ell)})$ typically first decreases and then increases with noticeable fluctuations, as shown in Fig.~\ref{Interfer_with_DeNum_Multiuser}. This behavior arises from the complex phase interactions among multi-user interference components. This behavior necessitates evaluating all possible candidate subsets to obtain the global minimum interference level.
\end{remark}

\subsubsection{Power allocation}
After the above dedicated antenna selection via the proposed method, the optimization problem \textbf{(P1)} reduces to optimizing the transmit power allocation $\{P_k\}$ to maximize the sum-rate
\begin{align}
	\textbf{(P10)}\;\; \max_{\{P_k\}}  \;\; \sum_{k=1}^K R^{\rm AS}_k \quad
	\text{s.t.}\;\; \eqref{TotPower}.
\end{align}
Problem \textbf{(P10)} can be efficiently solved by using the Lagrangian method or by following the same optimization procedure as in Problem \textbf{(P9)}, with details omitted for brevity.

\begin{remark}\rm (Computational complexity)
	For the proposed low-complexity algorithm, its computational complexities in the antenna selection and power allocation designs are in the orders of $O(N^2K^2)$ and $O((3K)^{3.5})$, respectively. 
	As a result, the overall computational complexity is dominated by the antenna selection procedure, since it scales with the number of antennas. Compared with the PDD-based method, the proposed low-complexity algorithm achieves a significantly lower computational complexity while maintaining comparable performance (see numerical results in Section~VI).  
\end{remark}

\vspace{-10pt}
\section{Numerical Results}
In this section, we present numerical results to evaluate the performance of the proposed antenna-selection-based flexible beamforming design and verify the effectiveness of  the developed algorithms. We consider a mixed-field communication system where a BS equipped with $N = 256$ antennas operates at $f = 30$~GHz, corresponding to an effective Rayleigh distance of 120~m at a spatial angle of $\theta = 0$ rad. The reference channel gain power is set as $\beta = (\lambda/4\pi)^2 = -62$~dB, and the noise power is $\sigma^2 = -80$~dBm. Unless otherwise specified, other parameters are set as $\rho = 800$, $c = 0.6$, and $\epsilon = 10^{-3}$.

For performance comparison, we consider the following baseline schemes (all based on our considered antenna-selection hardware architecture) and baseline architectures. 
\begin{itemize}
		
	\item \textit{Baseline scheme 1 (PDD-AS with equal power):} 
	The antenna selection is designed using the proposed PDD-based method, while the total transmit power is equally allocated among all users.
	
	\item \textit{Baseline scheme 2 (Random AS with optimized power):} For each user, antenna subsets are randomly generated over 2000 Monte Carlo trials. For each subset, the power allocation is optimized, and the subset achieving the maximum sum-rate is selected.
	
	\item \textit{Baseline architecture 1 (Common AS):} 
	A common antenna subset is selected and shared by all users, i.e., the same set of activated antennas is connected to all RF chains through a switch network.

	\item \textit{Baseline architecture 2 (Subarray selection):} 
	The BS antenna array is partitioned into $U$ non-overlapping subarrays, such that each user is served by at most one subarray.
	
	\item \textit{Baseline architecture 3 (Full-array w/o AS):} 
	A conventional fully-connected full-array architecture is considered, where MRT-based analog beamforming is employed and the transmit power allocation is optimized.
\end{itemize}

\begin{figure}[t]
	\centering
	\includegraphics[width=0.75\linewidth]{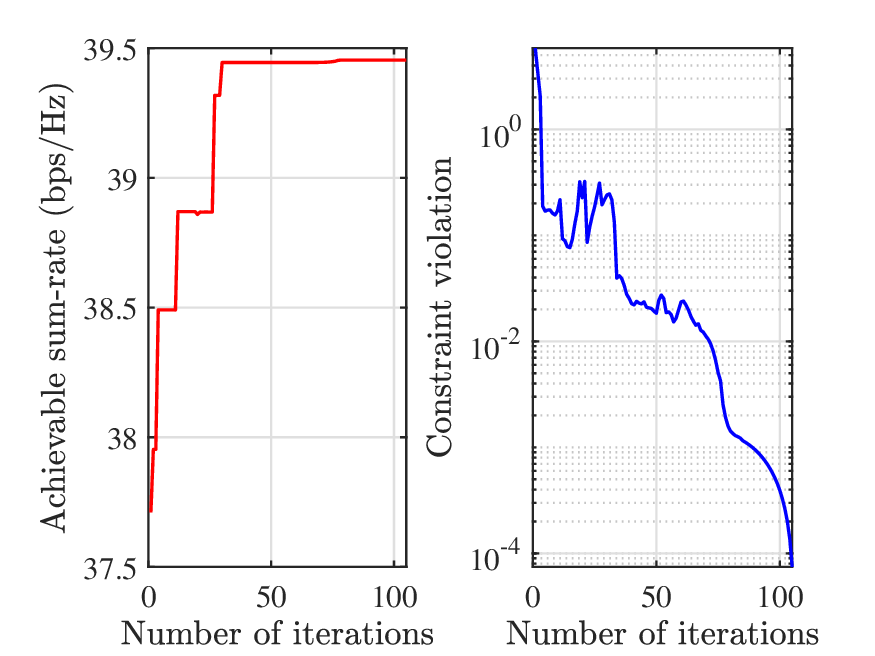}
	\caption{Convergence performance of proposed PDD-based algorithm.}
	\label{Convergence}
%	\vspace{-5pt}
\end{figure}

\vspace{-5pt}
\subsection{Convergence Performance}
We first evaluate the convergence performance of the proposed PDD-based algorithm in Fig. \ref{Convergence}, where 2 near-field users are located at (-0.30~rad, 5.82~m) and (0.17~rad, 6.21~m), and 3 far-field users are located at (-0.22~rad, 150~m), (0.09~rad, 175~m), (0.25~rad, 200~m), respectively. It is observed that the achievable sum-rate converges rapidly within less than 100 outer iterations, demonstrating high efficiency of the proposed PDD-based scheme. Moreover, the constraint violation $g$ reduces to a threshold of $10^{-4}$, confirming that the equality constraints in \eqref{P3SumNumAnten}, \eqref{IntroVhatBinary} and \eqref{IntroVhatEqual} are effectively satisfied.

\vspace{-5pt}
\subsection{Effect of Transmit Power}
\begin{figure}[t]
	\centering
	\includegraphics[width=0.75\linewidth]{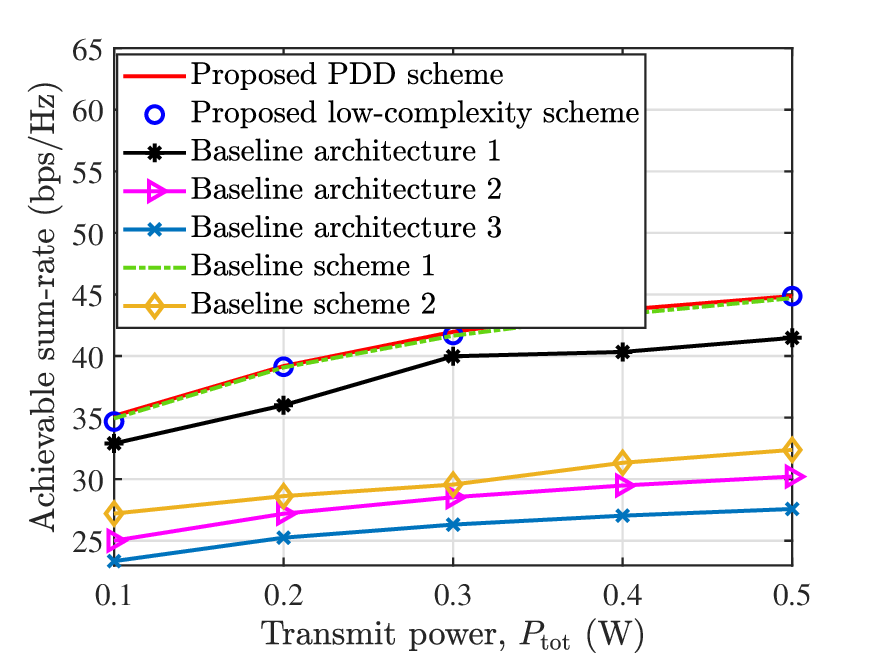}%{system_model_mixed_field.eps}
	\caption{Effect of the total transmit power.}
	\label{Rate_with_power}
	\vspace{-8pt}
\end{figure}

\begin{figure}[t]
	\centering
	\includegraphics[width=0.75\linewidth]{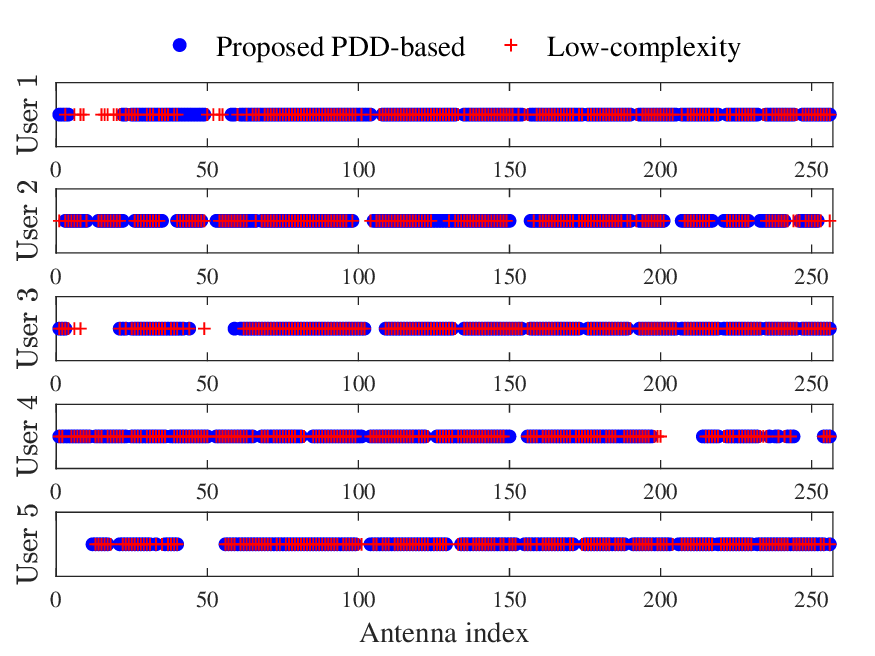}%{system_model_mixed_field.eps}
	\caption{Antenna activation patterns obtained by proposed PDD-based scheme and low-complexity scheme.}
	\label{Activation_status}
	\vspace{-5pt}
\end{figure}

In Fig.~\ref{Rate_with_power}, we plot the sum-rate performance of different schemes versus the total transmit power, under the same user location setup as in Fig.~\ref{Convergence}. It is observed that the proposed PDD-based scheme consistently achieves the best sum-rate performance across different transmit powers. Meanwhile, the proposed low-complexity greedy deactivation scheme achieves a sum-rate performance close to that of the PDD-based scheme. Compared with the baseline schemes, the proposed schemes achieve a significantly higher sum-rate than baseline scheme~2. In addition, the relatively small performance gap between the proposed schemes and baseline scheme~1 suggests that the sum-rate improvement is mainly attributed to the effective antenna selection, which suppresses most of the inter-user interference. As a result, \emph{far-field users can still be allocated with considerable transmit power}. Furthermore, in terms of hardware architectures, our proposed antenna-selection-based schemes significantly outperform the baseline architecture~3, as antenna selection allows the transmit beams to be flexibly reconstructed to suppress inter-user interference. Compared with the baseline architecture~2, our proposed schemes achieve a higher array gain because our proposed hardware architecture allows more flexible utilization of available antennas rather than dividing them into fixed subarrays. Furthermore, compared with the baseline architecture~1, which enforces a common antenna subset for all users, our schemes allow user-specific antenna selection.
\begin{table}[!t]
	\centering
	\caption{Runtime comparison.}
	\begin{tabular}{l|ccc}
		\hline
		\hline
		& \multicolumn{3}{c}{Runtime (s)} \\
		\cline{2-4}
		& $K=4$ & $K=5$ & $K=6$ \\
		\hline
		Proposed PDD-based scheme      & 544  & 727 & 939 \\
		\hline
		Proposed low-complexity scheme & 2.1    & 2.7    & 3.1 \\
		\hline
	\end{tabular}
\end{table}

To further verify the effectiveness of the proposed low-complexity greedy scheme, we present in Fig.~\ref{Activation_status} the antenna activation patterns obtained by the proposed PDD-based scheme and the low-complexity scheme, while we show in Table~I  their runtime for different numbers of users at $P_{\rm tot}=1$ W. As shown in Fig.~\ref{Activation_status}, the antenna activation patterns obtained by the low-complexity scheme are largely consistent with those of the PDD-based scheme, indicating that the former effectively captures the key antenna activation patterns for interference suppression. Moreover, it is observed in Table~I that the proposed low-complexity scheme achieves substantially shorter runtime than the PDD-based scheme, highlighting its superior computational efficiency.

\vspace{-5pt}
\subsection{Effect of Number of Near-field and Far-field Users}
\begin{figure}[t]
	\centering
	\begin{subfigure}[b]{0.75\linewidth}
		\includegraphics[width=1\linewidth]{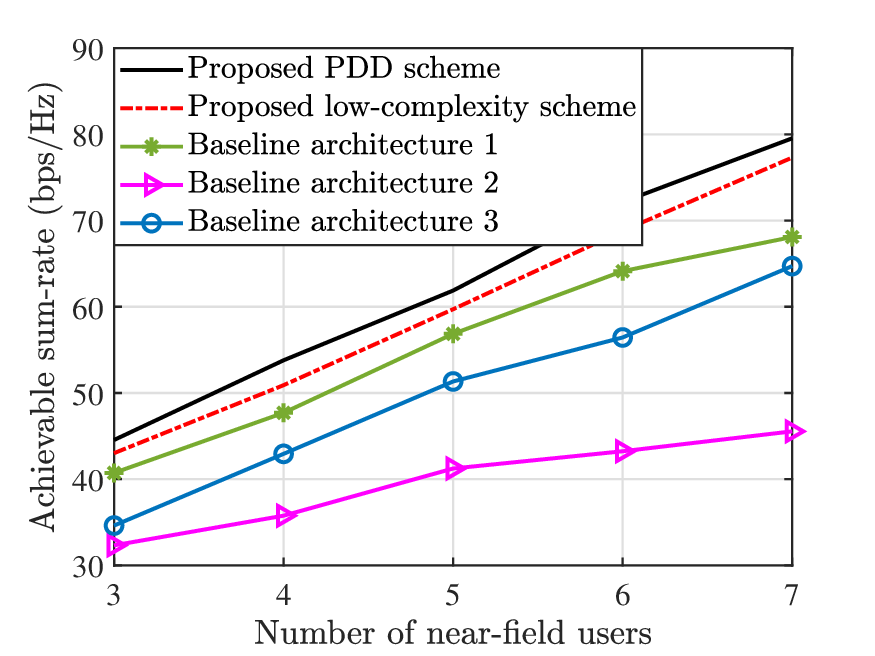}
		\caption{Achievable sum-rate vs. number of near-field users.}
		\label{Rate_with_NumNF}
	\end{subfigure}
	\begin{subfigure}[b]{0.75\linewidth}
		\includegraphics[width=1\linewidth]{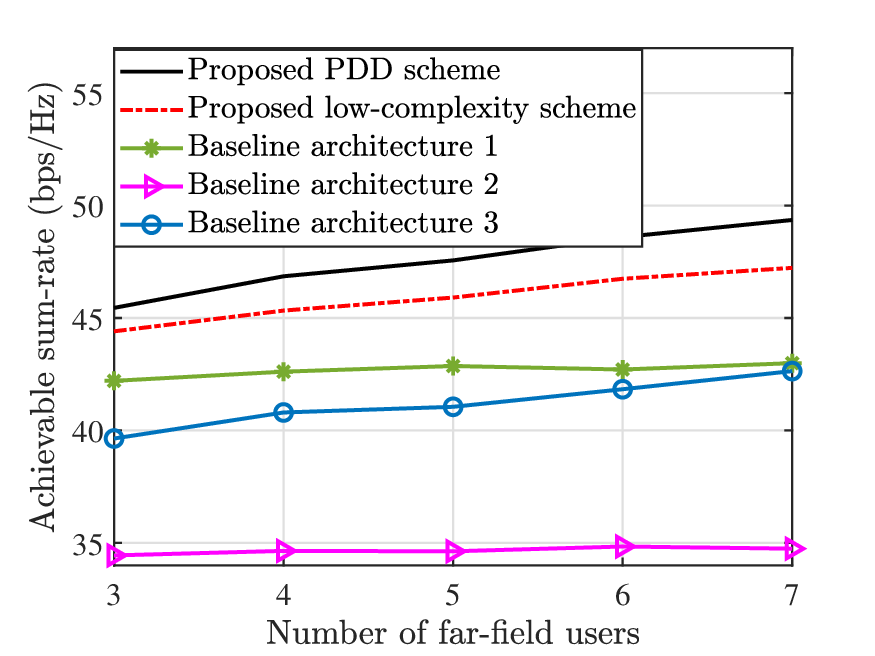}
		\caption{Achievable sum-rate vs. number of far-field users.}
		\label{Rate_with_NumFF}
	\end{subfigure}
	\caption{Effect of number of users.}
	\label{Rate_with_Numuser}
\end{figure}

In Fig.~\ref{Rate_with_Numuser}, we evaluate the achievable sum-rate performance of the proposed schemes under different numbers of users. Specifically, in Figs.~\ref{Rate_with_Numuser}(a) and~\ref{Rate_with_Numuser}(b), additional near-field and far-field users are randomly deployed within $r_k \in [0.05Z,0.2Z], k\in\mathcal{K}_1$ and $r_k \in [Z,2Z], k\in\mathcal{K}_2$, respectively, with $\theta_k \in [-\pi/3,\pi/3], k\in\mathcal{K}$.
First, it is observed from Fig. \ref{Rate_with_Numuser}(a) that the sum-rate increases significantly with the number of near-field users.
This is because the proposed scheme effectively mitigates multi-user interference in mixed-field environments, allowing the system to accommodate more users without compromising individual performance.
Furthermore, one can observe from Fig.~\ref{Rate_with_Numuser}(b) that the sum-rate also increases with the number of far-field users, albeit with a slower growth rate than that shown in Fig.~\ref{Rate_with_Numuser}(a).

\vspace{-5pt}
\subsection{Robustness of Proposed Schemes}
%\subsection{Effect of the Rician factor}
\begin{figure}[t]
	\centering
	\includegraphics[width=0.75\linewidth]{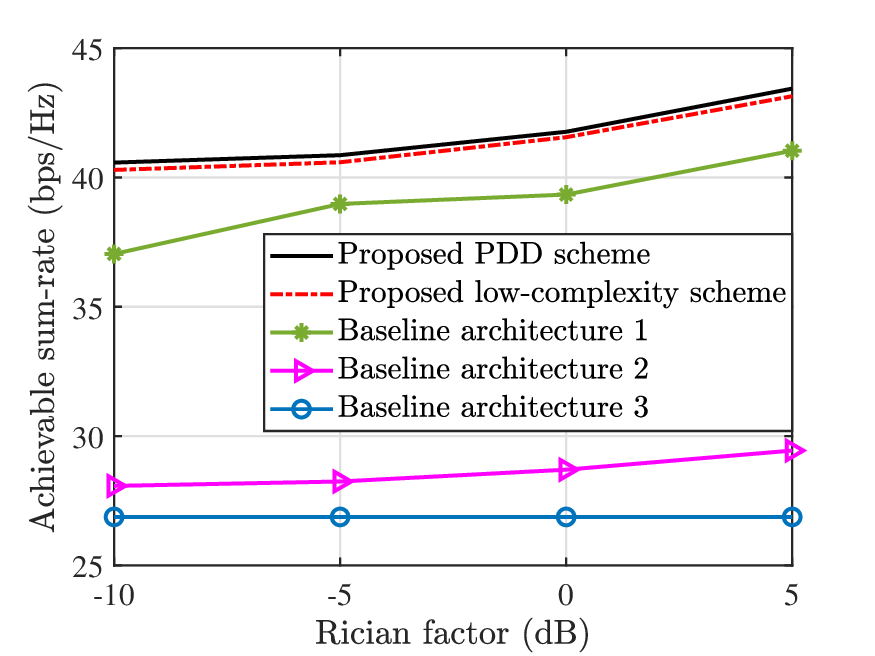}%{system_model_mixed_field.eps}
	\caption{Effect of Rician factors.}
	\label{Rate_with_Ricianfactor}
	\vspace{-15pt}
\end{figure}

\begin{figure}[t]
	\centering
	\includegraphics[width=0.75\linewidth]{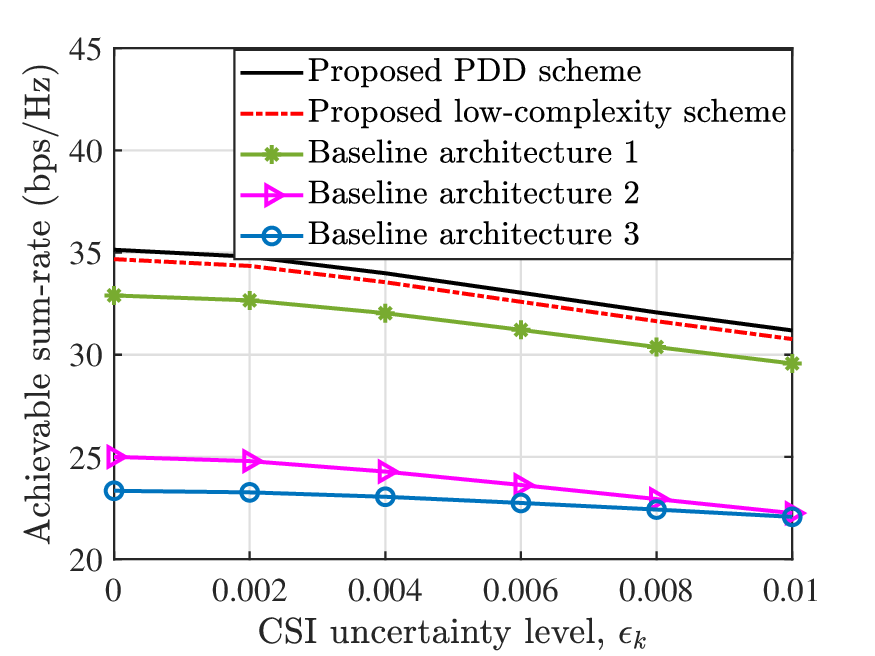}%{system_model_mixed_field.eps}
	\caption{Effect of CSI errors.}
	\label{Rate_with_error}
\end{figure}

\begin{figure}[t]
	\centering
	\includegraphics[width=0.75\linewidth]{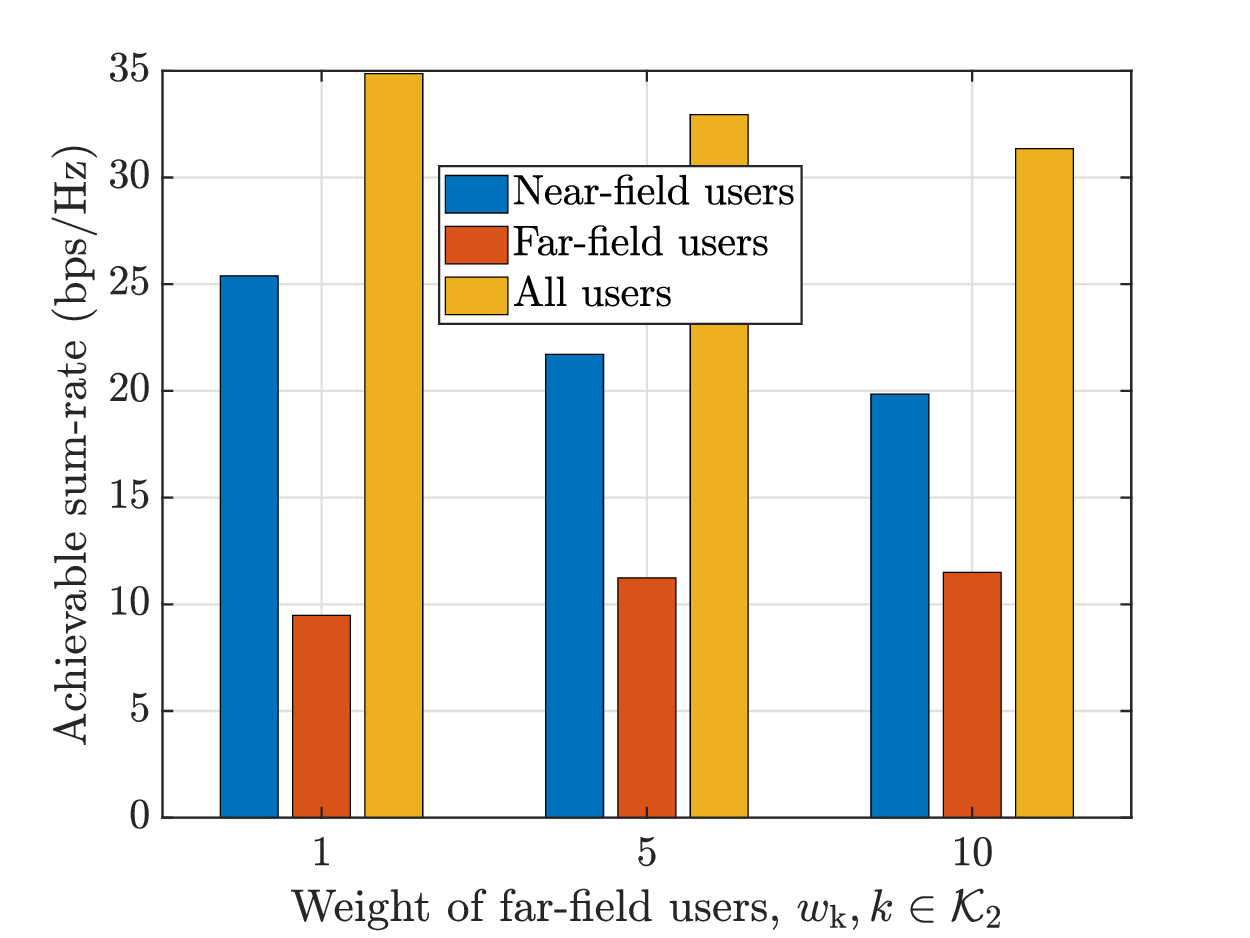}%{system_model_mixed_field.eps}
	\caption{Effect of weighting factors of far-field users.}
	\label{Rate_with_weight}
\end{figure}

First, we evaluate the sum-rate performance of proposed schemes under different Rician factors, as shown in Fig.~\ref{Rate_with_Ricianfactor}. It is observed that, even under a small Rician factor (e.g., $-10$ dB), both the proposed PDD-based scheme and the low-complexity scheme achieve a higher sum-rate than the baseline architectures. These results demonstrate the robustness and effectiveness of the proposed schemes in practical multi-path propagation environments.

Next, we investigate the impact of imperfect CSI on the achievable sum-rate performance. Specifically, the actual channel of user $k$ is modeled as $\mathbf{h}^H_k = \hat{\mathbf{h}}^H_k+\Delta\mathbf{h}^H_k,  \forall k \in \mathcal{K}$,
where $\hat{\mathbf{h}}^H_k$ denotes the estimated channel used for antenna selection design and $\Delta \mathbf{h}^H_k$ represents the channel estimation error. The error term is assumed to follow a complex Gaussian distribution, i.e., $\Delta\mathbf{h}^H_k\sim\mathcal{CN}(0,\mathbf{\Sigma}_k)$, with $\mathbf{\Sigma}_k=\epsilon^2_k\|\hat{\mathbf{h}}^H_k\|^2_2\mathbf{I}_N$, where $\epsilon_k$ denotes the CSI uncertainty level~\cite{Lyu_Bin2023}. We present in Fig.~\ref{Rate_with_error} the sum-rate achieved by different schemes versus the CSI uncertainty level. It is observed that the proposed schemes consistently outperform the baseline architectures in terms of sum-rate. Moreover, the sum-rate of proposed schemes exhibits only a moderate degradation as the CSI uncertainty increases, indicating the robustness of proposed algorithms against channel estimation errors.

Last, we extend Problem \textbf{(P2)} to a weighted sum-rate maximization to investigate the impact of user priority, formulated as $\max_{\{\mathbf{V}_k,P_k\}}\;\sum_{k \in \mathcal{K}_1} w_k R_k^{\rm AS}+\sum_{k \in \mathcal{K}_2} w_k R_k^{\rm AS}$, subject to \eqref{P2AntennaSelection} and \eqref{TotPower}, where $w_k$ denotes the priority weight of user $k$. This problem can be efficiently solved using the proposed PDD-based algorithm. With $w_k = 1$ for all near-field users ($\forall k \in \mathcal{K}_1$), we present in Fig.~\ref{Rate_with_weight} the sum-rates of near-field users, far-field users, and all users under different weights assigned to far-field users ($\forall k \in \mathcal{K}_2$). It is observed that increasing the far-field user weights improves their achievable rates at the expense of a moderate reduction in those of near-field users, while the sum-rate of all users degrades marginally. This demonstrates that the proposed scheme can flexibly optimize antenna selection and power allocation to achieve a favorable tradeoff between user fairness and overall system efficiency.

\vspace{-10pt}
\section{Conclusions}
In this paper, we considered an antenna-selection-based beamforming architecture for mixed-field multi-user communications, where a switch network dynamically connects each RF chain to a subset of PSs, enabling adaptive antenna selection and flexible beamforming. Based on this architecture, a joint antenna-selection and power-allocation problem is formulated to maximize the achievable sum-rate, which is non-convex and NP-hard.
To shed important insights, we first focused on a typical mixed-field two-user scenario and developed a low-complexity antenna deactivation algorithm to enhance near-far channel orthogonality by reshaping the array geometry. It is demonstrated that by strategically deactivating a small portion of antennas, the spatial orthogonality between near-field and far-field users can be significantly promoted, thereby achieving effective inter-user interference suppression.
For the general multi-user case, an efficient PDD-based two-layer optimization method was developed to obtain its high-quality solution. To further reduce computational complexity, a low-complexity antenna deactivation algorithm was also proposed for multi-user scenarios.
Numerical results demonstrated that the proposed schemes achieve substantial sum-rate improvement over various baseline schemes.

\vspace{-10pt}
\appendix

\vspace{-5pt}
\subsection{Proof for Lemma \ref{lemma_I_dff}} \label{I_diff}

Let $n^{(\ell+1)}$ denote the antenna removed at step $(\ell+1)$, such that $s^{(\ell+1)} = s^{(\ell)} - c_{n^{(\ell+1)}}$. In mixed-field XL-array system, when $\ell \ll N$, we have $|c_{n^{(\ell+1)}}| = 1/N \ll |s^{(\ell)}|$. Applying the first-order Taylor expansion $\sqrt{1+x} \approx 1+x/2$, the amplitude of $s^{(\ell+1)}$ is approximated as
{\small
	\begin{equation}
		|s^{(\ell+1)}| = |s^{(\ell)}| \sqrt{1\!+\! \frac{1}{N^2|s^{(\ell)}|^2}\!-\!\frac{2 \cos\theta_{\Delta}^{(\ell+1)}}{N|s^{(\ell)}|}} \!\approx |s^{(\ell)}| - \frac{\cos\theta_{\Delta}^{(\ell+1)}}{N}, \nonumber
\end{equation}}where $\theta_{\Delta}^{(\ell+1)} \triangleq \angle c_{n^{(\ell+1)}} - \angle s^{(\ell)}$. The reduction in the interference coupling factor, denoted by $I^{(\ell+1)}_{\Delta}$, is given by
{\small
	\begin{align}
		I^{(\ell+1)}_{\Delta} &\approx \left( \frac{N}{\sqrt{N-\ell}} - \frac{N}{\sqrt{N-\ell-1}} \right) |s^{(\ell)}| + \frac{\cos\theta_{\Delta}^{(\ell+1)}}{\sqrt{N-\ell-1}}. \label{ReduceInterf_proof}
\end{align}}Utilizing the approximation $1/\sqrt{(N-\ell-1)} \approx 1/\sqrt{(N-\ell)}$ for $\ell \ll N$, the first term in \eqref{ReduceInterf_proof} vanishes, yielding $I^{(\ell+1)}_{\Delta} \approx \cos(\theta_{\Delta}^{(\ell+1)})/\sqrt{N-\ell}$, thus completing the proof.

\vspace{-5pt}
\subsection{Proof for Lemma \ref{Lemma_optimal_number}} \label{proof_optimal_number}
According to \eqref{AchinforRate_approx_uN1} and \eqref{AchinforRate_approx_uF1}, the objective function of Problem~\textbf{(P3.1)} can be approximated as
\begin{align}
	\tilde{R}^{\rm AS}_1+\tilde{R}^{\rm AS}_2 = \log_2\left(\bar{\Gamma}_1(\ell_1) \cdot \bar{\Gamma}_2(\ell_2)\right),
\end{align}  
where $\bar{\Gamma}_1(\ell_1)$ and $\bar{\Gamma}_1(\ell_1)$ are respectively given by
{\small
\begin{align}
	&\bar{\Gamma}_1(\ell_1) \triangleq \frac{P_1|h_1|^2\!\left(N-\ell_1\right)} {P_1|h_2|^2\!\left(I_1^{(0)} - \alpha_1 \ell_1\right)^2+\sigma^2},\nonumber\\
	&\bar{\Gamma}_2(\ell_2) \triangleq \frac{P_2|h_2|^2\!\left(N-\ell_2\right)} {P_2|h_1|^2\!\left(I_2^{(0)} - \alpha_2 \ell_2\right)^2+\sigma^2}.
\end{align}}
Since each term depends only on the antenna deactivation variable of a single user, $\ell_1$ and $\ell_2$ can be optimized independently. Without loss of generality, we first consider the antenna selection for the near-field user and aim to maximize $\bar{\Gamma}_1$ with respect to $\ell_1$.
Define $A\triangleq P_1|h_1|^2\!\left(N-\ell_1\right)$ and $B\triangleq P_1|h_2|^2\!\left(I_1^{(0)} - \alpha_1 \ell_1\right)^2+\sigma^2.$
Then, we have $\bar{\Gamma}_1=A/B$, whose first-order derivative with respect to $\ell_1$ is given by
\begin{align}
	\frac{\partial \bar{\Gamma}_1}{\partial \ell_1}=\left(	\frac{\partial A}{\partial \ell_1}B-\frac{\partial B}{\partial \ell_1}A\right)/B^2. \label{derivative}
\end{align}
The monotonicity of $\bar{\Gamma}_1$ is determined by the numerator of \eqref{derivative}, which is a quadratic function of $\ell_1$. By setting this numerator to zero, we obtain a closed-form solution for the number of deactivated antennas for the near-field user, as given in \eqref{Num_Opt_NF}. Following the same procedure, the solution for the far-field user in \eqref{Num_Opt_FF} can be obtained, which completes the proof.

\vspace{-5pt}

\bibliographystyle{IEEEtran}
\bibliography{Refs}

\end{document}